\documentclass[ALICE,manyauthors]{cernphprep}

\usepackage[comma,square,numbers,sort&compress]{natbib}
\usepackage{hyperref}
\usepackage{lineno}
\usepackage{color}
\usepackage{floatrow}
\usepackage{verbatim}
\usepackage[T1]{fontenc}
\usepackage{orcidlink}
\begin{document}

%%%%%%%%%%%%%%%%%%%%%%%%%%%%%%%%%%%%%%%%%%%%%%%%%%
% These are some new commands that may be useful 
% for paper writing in general. If other newcommands
% are needed for your specific paper, please feel 
% free to add here. 
%
% The currently available commands are organized in: 
% 1) Systems
% 2) Quantities
% 3) Energies and units
% 4) Detectors
% 5) particle species 
%%%%%%%%%%%%%%%%%%%%%%%%%%%%%%%%%%%%%%%%%%%%%%%%%%

% 1) SYSTEMS 
\newcommand{\pp}           {pp\xspace}
\newcommand{\ppbar}        {\mbox{$\mathrm {p\overline{p}}$}\xspace}
\newcommand{\XeXe}         {\mbox{Xe--Xe}\xspace}
\newcommand{\PbPb}         {\mbox{Pb--Pb}\xspace}
\newcommand{\pPb}          {\mbox{p--Pb}\xspace}
\newcommand{\AuAu}         {\mbox{Au--Au}\xspace}
\newcommand{\dAu}          {\mbox{d--Au}\xspace}

% 2) QUANTITIES 
\newcommand{\roots}        {\ensuremath{\sqrt{s}}\xspace}
\newcommand{\snn}          {\ensuremath{\sqrt{s_{\mathrm{NN}}}}\xspace}
\newcommand{\pt}           {\ensuremath{p_{\rm T}}\xspace}
\newcommand{\meanpt}       {$\langle p_{\mathrm{T}}\rangle$\xspace}
\newcommand{\ycms}         {\ensuremath{y_{\rm CMS}}\xspace}
\newcommand{\ylab}         {\ensuremath{y_{\rm lab}}\xspace}
\newcommand{\etarange}[1]  {\mbox{$\left | \eta \right |~<~#1$}}
\newcommand{\yrange}[1]    {\mbox{$\left | y \right |~<~#1$}}
\newcommand{\dndy}         {\ensuremath{\mathrm{d}N_\mathrm{ch}/\mathrm{d}y}\xspace}
\newcommand{\dndeta}       {\ensuremath{\mathrm{d}N_\mathrm{ch}/\mathrm{d}\eta}\xspace}
\newcommand{\avdndeta}     {\ensuremath{\langle\dndeta\rangle}\xspace}
\newcommand{\dNdy}         {\ensuremath{\mathrm{d}N_\mathrm{ch}/\mathrm{d}y}\xspace}
\newcommand{\Npart}        {\ensuremath{N_\mathrm{part}}\xspace}
\newcommand{\Ncoll}        {\ensuremath{N_\mathrm{coll}}\xspace}
\newcommand{\dEdx}         {\ensuremath{\textrm{d}E/\textrm{d}x}\xspace}
\newcommand{\RpPb}         {\ensuremath{R_{\rm pPb}}\xspace}
\newcommand{\kstar}        {\ensuremath{k^*}\xspace}
\newcommand{\mt}           {\ensuremath{m_{\rm{T}}}\xspace}

% 3) ENERGIES, UNITS
\newcommand{\nineH}        {$\sqrt{s}~=~0.9$~Te\kern-.1emV\xspace}
\newcommand{\seven}        {$\sqrt{s}~=~7$~Te\kern-.1emV\xspace}
\newcommand{\thirteen}     {$\sqrt{s}~=~13$~Te\kern-.1emV\xspace}
\newcommand{\forteen}     {$\sqrt{s}~=~14$~Te\kern-.1emV\xspace}
\newcommand{\twoH}         {$\sqrt{s}~=~0.2$~Te\kern-.1emV\xspace}
\newcommand{\twosevensix}  {$\sqrt{s}~=~2.76$~Te\kern-.1emV\xspace}
\newcommand{\five}         {$\sqrt{s}~=~5.02$~Te\kern-.1emV\xspace}
\newcommand{\fivefive}     {$\sqrt{s}~=~5.5$~Te\kern-.1emV\xspace}
\newcommand{\twosevensixnn}{$\sqrt{s_{\mathrm{NN}}}~=~2.76$~Te\kern-.1emV\xspace}
\newcommand{\fivenn}       {$\sqrt{s_{\mathrm{NN}}}~=~5.02$~Te\kern-.1emV\xspace}
\newcommand{\LT}           {L{\'e}vy-Tsallis\xspace}
\newcommand{\tev}          {TeV\xspace}
\newcommand{\gev}          {GeV\xspace}
\newcommand{\mev}          {MeV\xspace}
\newcommand{\GeVmass}      {GeV/$c^2$\xspace}
\newcommand{\MeVmass}      {MeV/$c^2$\xspace}
\newcommand{\lumi}         {\ensuremath{\mathcal{L}}\xspace}
% 3) ENERGIES, UNITS v2
\newcommand{\onethree}        {$\sqrt{s}~=~13$~Te\kern-.1emV\xspace}
\newcommand{\MeV}  {\ensuremath{\text{Me\kern-.1emV}}\xspace}
\newcommand{\MeVc}  {\ensuremath{\text{Me\kern-.1emV/}c}\xspace}
\newcommand{\MeVcc}  {\ensuremath{\text{Me\kern-.2emV/}c^2}\xspace}
\newcommand{\GeV}  {\ensuremath{\text{Ge\kern-.1emV}}\xspace}
\newcommand{\GeVc}  {\ensuremath{\text{Ge\kern-.1emV/}c}\xspace}
\newcommand{\GeVcc}  {\ensuremath{\text{Ge\kern-.2emV/}c^2}\xspace}
\newcommand{\TeV}  {\ensuremath{\text{Te\kern-.1emV}}\xspace}

% 4) DETECTORS 
\newcommand{\ITS}          {\rm{ITS}\xspace}
\newcommand{\TOF}          {\rm{TOF}\xspace}
\newcommand{\ZDC}          {\rm{ZDC}\xspace}
\newcommand{\ZDCs}         {\rm{ZDCs}\xspace}
\newcommand{\ZNA}          {\rm{ZNA}\xspace}
\newcommand{\ZNC}          {\rm{ZNC}\xspace}
\newcommand{\SPD}          {\rm{SPD}\xspace}
\newcommand{\SDD}          {\rm{SDD}\xspace}
\newcommand{\SSD}          {\rm{SSD}\xspace}
\newcommand{\TPC}          {\rm{TPC}\xspace}
\newcommand{\TRD}          {\rm{TRD}\xspace}
\newcommand{\TZERO}        {\rm{T0}\xspace}
\newcommand{\VZERO}        {\rm{V0}\xspace}
\newcommand{\VZEROA}       {\rm{V0A}\xspace}
\newcommand{\VZEROC}       {\rm{V0C}\xspace}
\newcommand{\Vdecay} 	   {\ensuremath{V^{0}}\xspace}

% 4) PARTICLE SPECIES 
\newcommand{\phot}         {\ensuremath{\gamma}\xspace} 
\newcommand{\ee}           {\ensuremath{\mathrm{e}^{+}\mathrm{e}^{-}}\xspace} 
\newcommand{\pip}          {\ensuremath{\uppi^{+}}\xspace}
\newcommand{\pim}          {\ensuremath{\uppi^{-}}\xspace}
\newcommand{\kap}          {\ensuremath{\rm{K}^{+}}\xspace}
\newcommand{\kam}          {\ensuremath{\rm{K}^{-}}\xspace}
\newcommand{\pbar}         {\ensuremath{\rm\overline{p}}\xspace}
\newcommand{\kzero}        {\ensuremath{{\rm K}^{0}_{\rm{S}}}\xspace}
\newcommand{\lmb}          {\ensuremath{\upLambda}\xspace}
\newcommand{\sigm}         {\ensuremath{\upSigma}\xspace}
\newcommand{\almb}         {\ensuremath{\overline{\upLambda}}\xspace}
\newcommand{\Om}           {\ensuremath{\upOmega^-}\xspace}
\newcommand{\Mo}           {\ensuremath{\overline{\upOmega}^+}\xspace}
\newcommand{\X}            {\ensuremath{\upXi^-}\xspace}
\newcommand{\Ix}           {\ensuremath{\overline{\upXi}^+}\xspace}
\newcommand{\Xis}          {\ensuremath{\upXi^{\pm}}\xspace}
\newcommand{\Oms}          {\ensuremath{\upOmega^{\pm}}\xspace}
\newcommand{\degr}         {\ensuremath{^{\rm o}}\xspace}

\newcommand{\siZ}          {\ensuremath{\upSigma^{0}}\xspace}
\newcommand{\siP}          {\ensuremath{\upSigma^{+}}\xspace}
\newcommand{\siM}          {\ensuremath{\upSigma^{-}}\xspace}
\newcommand{\asiZ}         {\ensuremath{\overline{\upSigma^{0}}}\xspace}
\newcommand{\psiZCF}       {\rm{p}\mbox{--}\siZ}
\newcommand{\SNCF}         {\rm{N}\mbox{--}\ensuremath{\upSigma}\xspace}
\newcommand{\LNCF}         {\rm{N}\mbox{--}\ensuremath{\upLambda}\xspace}
\newcommand{\apasiZCF}     {\ensuremath{\overline{\mathrm{p}}}\mbox{--}\asiZ}
\newcommand{\psiZCombCF}   {\rm{p}\mbox{--}\siZ \ensuremath{\oplus} \apasiZCF}
\newcommand{\SB}           {\ensuremath{(\upLambda\gamma)}\xspace}
\newcommand{\pSBCF}        {\rm{p}\mbox{--}\SB}

\newcommand{\pLCF}         {\rm{p}\mbox{--}\lmb}
\newcommand{\nSplusCF}     {\rm{n}\mbox{--}\ensuremath{\upSigma^{+}}}

\newcommand{\Ledn}         {Lednick\'y--Lyuboshits approach\xspace}
\newcommand{\chiEFT}       {\ensuremath{\chi}\rm{EFT}\xspace}
\newcommand{\fss}       {\rm{fss2}\xspace}
\newcommand{\ESC}          {\rm{ESC16}\xspace}
\newcommand{\NSC}          {\rm{NSC97f}\xspace}

\newcommand{\pP}           {\ensuremath{\mbox{p--p}}\xspace}
\newcommand{\ApAP}         {\ensuremath{\mbox{\pbar--\pbar}}\xspace}
\newcommand{\pPComb}       {\ensuremath{\mbox{p--p} \oplus \mbox{\pbar--\pbar}}\xspace}

\newcommand{\radiusResult} {\ensuremath{r_0 = 1.249 \pm 0.008\, \mathrm{(stat)} \,^{+0.024} _{-0.021}\, \mathrm{(syst)}}\,fm\xspace}

% 5) Three body femto
\newcommand{\pppCF}         {\rm{p}\mbox{--}\rm{p}\mbox{--}\rm{p}\xspace}
\newcommand{\ppLCF}         {\rm{p}\mbox{--}\rm{p}\mbox{--}\lmb}
\newcommand{\apapapCF}         {\pbar\mbox{--}\pbar\mbox{--}\pbar\xspace}
\newcommand{\apapaLCF}         {\pbar\mbox{--}\pbar\mbox{--}\almb}

\newcommand{\ppspCF}         {(\rm{p}\mbox{--}\rm{p})\mbox{--}\rm{p}\xspace}
\newcommand{\ppsLCF}         {(\rm{p}\mbox{--}\rm{p})\mbox{--}\lmb\xspace}
\newcommand{\ppLsCF}         {\rm{p}\mbox{--}(\rm{p}\mbox{--}\lmb)\xspace}

\newcommand{\pppbarCF}         {\rm{p}\mbox{--}\rm{p}\mbox{--}\pbar\xspace}

\newcommand{\ppbarCF}         {\rm{p}\mbox{--}\pbar\xspace}

\newcommand{\ppCF}         {\rm{p}\mbox{--}\rm{p}\xspace}

\newcommand{\ppspbarCF}         {(\rm{p}\mbox{--}\rm{p})\mbox{--}\pbar\xspace}

\newcommand{\pppbarsCF}         {(\rm{p}\mbox{--}\pbar)\mbox{--}\rm{p}\xspace}

\newcommand{\Qthree}           {\ensuremath{Q_3}\xspace}

% Comments
\newcommand{\ls}[1]{{\color{red}[LS: #1]}}

%%%%%%%%%%%%%%%  Title page %%%%%%%%%%%%%%%%%%%%%%%%
\begin{titlepage}
\PHyear{2022}
\PHnumber{110}      % required, will be obtained from PH
\PHdate{30 May}  % required, will be obtained from PH

\title{Towards the understanding of the genuine three-body interaction for p--p--p and p--p--$\Lambda$} %genuine three-baryon interaction}
\ShortTitle{Three-baryon interaction} 

\Collaboration{ALICE Collaboration\thanks{See Appendix~\ref{app:collab} for the list of collaboration members}}
\ShortAuthor{ALICE Collaboration} 

\begin{abstract}
Three-body nuclear forces play an important role in the structure of nuclei and hypernuclei and are also incorporated in models to describe the dynamics of dense baryonic matter, such as in neutron stars.
So far, only indirect measurements anchored to the binding energies of nuclei can be used to constrain the three-nucleon force, and if hyperons are considered, the scarce data on hypernuclei impose only weak constraints on the three-body forces. 
In this work, we present the first direct measurement of the \pppCF and \ppLCF systems in terms of three-particle correlation functions carried out for pp collisions at $\sqrt{s} = 13$ TeV. Three-particle cumulants are extracted from the correlation functions by applying the Kubo formalism, where the three-particle interaction contribution to these correlations can be isolated after subtracting the known two-body interaction terms. A negative cumulant is found for the \pppCF system, hinting to the presence of a residual three-body effect while for \ppLCF the cumulant is consistent with zero. This measurement demonstrates the accessibility of three-baryon correlations at the LHC.

\end{abstract}
\end{titlepage}
\setcounter{page}{2}

\section{Introduction}
One of the open challenges of nuclear physics is the understanding of many-particle dynamics.  
Studies of the nuclear structure have unambiguously shown that calculations based only on nucleon--nucleon (N--N) interactions fail to accurately describe many experimental observables, such as nuclear binding energies along the periodic table of elements~\cite{navratil2016unified}, the position of the neutron drip line for neutron-rich nuclei~\cite{doi:10.1146/annurev-nucl-102313-025446} or the properties of the recently observed four-neutrons resonance~\cite{Duer:2022ehf}.
A significant improvement in the modelling of nuclear bound objects has been achieved by including 
three-body forces in theoretical calculations. 
These three-body forces are implemented in chiral effective field theories~\cite{RevModPhys.81.1773} and in a number of ab initio many-body methods such as no-core shell model~\cite{barrett2013ab}, coupled-cluster theory~\cite{hagen2012evolution,hagen2014coupled}, self-consistent Green's function theory~\cite{soma2013ab}, similarity renormalisation group~\cite{roth2011similarity,stroberg2019nonempirical}, and quantum Monte Carlo~\cite{carlson2015quantum}.  
Studies conducted on intermediate mass neutron-rich nuclei proved that the sensitivity to the three-body forces increases  with the number of neutrons in the system~\cite{hagen2012evolution}. 
Three-body forces within light and medium-mass nuclei,
where the nuclear saturation density corresponds to typical inter-particle distances of 2 fm, contribute about $10-15$ \% to the total interaction strength~\cite{RevModPhys.85.197,Kievsky_2008}.
However, at higher densities and shorter inter-particle distances their contribution might increase~\cite{doi:10.1146/annurev-nucl-102313-025446}, but no data are available in such a regime and the properties of nuclear matter can be only extrapolated using the available information at saturation densities.
The experimental information on the three-body forces involving $\Lambda$ hyperons is even more scarce since the data available for hypernuclei are much less than the data for nuclei. Recent  hypetriton measurements in several colliding systems at RHIC and LHC~\cite{STAR:2017gxa,ALICE:2019vlx,STAR:2019wjm,ALargeIonColliderExperiment:2021puh} provide important input to the understanding of N--N--$\Lambda$ forces and future measurements will resolve the current tensions among the different estimations of the binding energy and life-time. Theoretical works assign to the hypertriton a radius of the order of 5 fm~\cite{10.1143/PTP.103.929} and hence a N--$\Lambda$ distance of 10 fm~\cite{PhysRevC.100.034002,Kievsky_2008} within this state. 
Heavier hypernuclei are more compact and their size is comparable to that of normal nuclei so that they represent an optimal test bed for the N--N--$\Lambda$ interaction~\cite{Feliciello:2015dua}. 
However, good fits of the theoretical models to the available hypernuclear data, from ${}_\Lambda^{7}$Li to ${}_\Lambda^{208}$Pb~\cite{PhysRevC.53.1210,agnello2011hypernuclear,PhysRevLett.120.132505,PhysRevC.99.054309}, require a full understanding of the shell-structure of such bound objects as well as accurate experimental constraints on the spin-dependent N--$\Lambda$ interaction, in particular for the p-wave and higher partial waves. The lack of precise data as well as the limitations in the microscopic description of the structure of hypernuclei cause large ambiguities on the strength of the N--N--$\Lambda$ three-body force. Further opportunities are provided by the recently observed $^{3}_{\Lambda}$n bound state~\cite{PhysRevC.88.041001} and planned experimental programs focused on neutron-rich hypernuclei~\cite{Saito:2021gao}.
Nevertheless, the contribution from three-body forces in bound objects such as nuclei and hypernuclei cannot be separated from the lower-order two-body interactions, hence, complementary experimental methods to investigate three-baryon systems could provide an important contribution to this field.

Neutron-rich and dense baryonic matter constitutes an interesting system also because of its connections to the physics of neutron stars (NS)~\cite{Oertel:2016bki}. 
The structure and composition of the innermost part of NS is not known. Amongst many possible scenarios, 
some models support the appearance of various hadronic particle species with increasing baryon density inside the star~\cite{Oertel:2016bki,Tolos:2020aln}. The presence of hadronic degrees of freedom and their relative abundances
are sensitive to the two- and three-body interaction models which are used to compute the equation of state (EoS) of NS matter. The different hypotheses can be tested by deriving the masses and radii of NS for a specific EoS and comparing them with the corresponding astrophysical observations~\cite{Tolos:2020aln}. The suggestion of strange baryons inside NS is motivated by the fact that central densities of NS might become sufficiently large ($\rho \approx 3-4 \rho_0$, where $\rho_0$ is the nuclear saturation density) to provide favourable conditions for the onset of strangeness production processes leading to, in particular, the formation of hyperons. The appearance of $\Lambda$ hyperons in NS matter results in a softening of the EoS which is at variance with astrophysical observations of two solar mass stars~\cite{natureNS,Antoniadis1233232}. 
However, in Ref.~\cite{Lonardoni:2014bwa}, it was shown that by adding a strongly repulsive N--N--$\Lambda$ interaction, tuned to reproduce the separation energies of $\Lambda$ hyperons in several hypernuclei, a sufficiently stiff EoS can be obtained and even the massive NS observables can be reproduced. This indicates that three-body forces may have a significant contribution in models that describe the structure of NS.
Hence, a direct measurement of the three-body forces involving nucleons and hyperons at small inter-particle distances is required.

The femtoscopy technique can be used as a tool to investigate the strong interaction amongst hadrons produced in particle collisions~\cite{wiedemann1999particle,heinz1999two,lednicky2004correlation,Lisa:2005dd} and recently has been successfully employed to analyse experimental data.
The produced hadrons may undergo final state interactions (FSIs) and the resulting correlation in the momentum space can be studied to test the underlying dynamics using correlation functions~\cite{lednicky2004correlation,Lisa:2005dd}. The method has been applied by the STAR Collaboration to measure hadron-hadron correlations in Au--Au collisions with a centre-of-mass energy of 200 GeV per nucleon pair~\cite{adamczyk2015lambda,STAR:2015kha,STAR:2018uho}. In such ultra-relativistic heavy-ion collisions, the average relative distances of emitted particles is about 7-8 fm~\cite{Lisa:2005dd}. In small colliding systems, such as pp and p--Pb collisions at the LHC, particles are produced at distances of the order of 1 fm, hence, the sensitivity of the correlation function to the short-range strong interaction is enhanced. 
Recently, the method has been employed by ALICE to study FSIs of hadrons produced in such small colliding systems. The large data samples allowed for the precise measurement of correlation functions for multiple hadronic pairs (p--p~\cite{ALICE:Run1}, p--K$^+$ and p--K$^-$~\cite{Acharya:2019bsa}, p--$\mathrm{\Lambda}$~\cite{ALICE:Run1}, p--$\mathrm{\Sigma}^0$~\cite{ALICE:pSig0}, $\mathrm{\Lambda}$--$\mathrm{\Lambda}$~\cite{ALICE:LL}, p--$\mathrm{\Xi}^-$~\cite{ALICE:pXi}, p--$\mathrm{\Omega}^-$~\cite{ALICE:pOmega}, p--$\phi$~\cite{ALICE:2021cpv} and baryon--antibaryon~\cite{ALICE:2021cyj}). By using these results, several models for the two-body strong interaction could be validated (for a complete review see Ref.~\cite{Fabbietti:2020bfg}). 

The femtoscopy technique was also employed in the analysis of 
three- and four-pion correlations measured in pp, p--Pb, Pb--Pb collision systems by ALICE~\cite{Dhevan,PhysRevC.93.054908} to probe coherent hadron production. The Kubo's cumulant expansion method~\cite{Kubo} was used to isolate the genuine three-particle correlation from the two-body contributions where the latter were evaluated by combining two particles from the same event and a third particle taken from another event.
Alternatively, the recently developed projector method~\cite{DelGrande:2021mju}, where either the theoretical or the measured two-body correlation functions are used to obtain the lower-order contributions, can be employed. This method allows a significant reduction of the statistical uncertainties.

In this article, the first femtoscopic study of three-baryon correlations is performed for the \pppCF and \ppLCF systems measured in high-multiplicity (HM) pp collisions at $\sqrt{s} = 13$ TeV. The Kubo's formalism and the projector method are employed to isolate the genuine three-body correlation and the choice of the reaction system aims to study the interaction at small distances.
The article is organised as follows: in Section~\ref{sec:eventselection} the data analysis procedure is presented starting from the event selection; in Section~\ref{3corr} the definition of the two-particle correlation function is extended to the three-particle case; in Section~\ref{sec:cumulants} the femtoscopic three-particle cumulant is defined; the lower-order two-particle correlation contributions in the measured correlation functions are evaluated in Section~\ref{sec:projector} and the decomposition of the cumulant to account for misidentifications and particle feed-down are presented in Section~\ref{sec:decomposition}; the final results are discussed in Section~\ref{sec:results} and the conclusions are given in Section~\ref{sec:conclusions}.

\section{Analysis}
\label{sec:analysis}
\subsection{Event selection and particle identification}
\label{sec:eventselection}
The data sample of \pp collisions at a centre of mass energy \thirteen was recorded with the ALICE detector~\cite{ALICE, ALICEperf} during the LHC Run 2 (2015--2018). The sample has been collected employing a HM trigger.
The trigger is based on the measured amplitude in the V0 detector system, consisting of two arrays of plastic scintillators located at forward ($2.8<\eta<5.1$) and backward ($-3.7<\eta<-1.7$) pseudorapidities~\cite{Abbas:2013taa}. The selected HM events correspond to the highest 0.17\%  multiplicity interval with respect to all  inelastic collisions with at least one measured charged particle within $|\eta|<1$ (INEL$>0$). This condition results in an average of 30 charged particles in the range $|\eta|<0.5$~\cite{ALICE:pOmega}.
Charged-particle tracking in the midrapidity region is conducted with the Inner Tracking System (\ITS)~\cite{ALICE} and the Time Projection Chamber (\TPC)~\cite{TPC}. These detectors are immersed in a homogeneous 0.5 T magnetic field parallel to the beam direction.
The \ITS consists of six cylindrical layers of high position-resolution silicon detectors placed radially between $3.9$ and 43 cm around the beam vacuum tube.
The \TPC consists of a 5 m long, cylindrical gaseous detector with full azimuthal coverage in the pseudorapidity range $|\eta| < 0.9$.

Particle identification (PID) is conducted via the measurement of the specific ionisation energy loss (\dEdx) in the TPC gas with up to 159 reconstructed space points along the particle trajectory. For high momentum particles, the TPC measurement is combined with information provided by the Time-Of-Flight (\TOF)~\cite{TOF} detector system, which is located at a radial distance of 3.7~m from the nominal interaction point and consists of Multigap Resistive Plate Chambers covering the full azimuthal angle in $|\eta|<0.9$.

The primary vertex (PV) of the event is reconstructed with the combined track information of the \ITS and the \TPC, and independently with track segments in the two innermost layers of the ITS. 
The reconstructed PV of the event is required to have a maximal displacement with respect to the nominal interaction point of 10~cm along the beam axis, in order to ensure a uniform acceptance. Pile-up events with multiple primary vertices are removed following the procedure described in Refs.~\cite{ALICE:Run1,ALICE:pXi,Acharya:2020dfb}. This rejects the events with pile-up of collisions occurring in the same or nearby bunch crossings. However, additional clean-up has to be applied on the track selection level to reject particles produced in pile-up collisions in the long TPC readout time.

 A total of $1.0\times 10 ^9$ HM events are used for the analysis after event selection. In order to build the three-particle correlation functions of \pppCF and \ppLCF systems, particle and antiparticle distributions are combined. 
 In the following, \pppCF refers to \pppCF $\oplus$ \apapapCF and \ppLCF refers to \ppLCF $\oplus$ \apapaLCF.
The proton and \lmb candidates as well as their antiparticles need to be selected. As the particle and antiparticle selections are identical, only the particles are explicitly discussed below. Both particle species are reconstructed using the procedure described in Ref.~\cite{Acharya:2020dfb}, while the related systematic uncertainties are evaluated by varying the kinematic and topological selection criteria used in the reconstruction. In the following text, the systematic variations are enclosed in parentheses. 
 
 The primary protons are selected in the momentum interval $0.5~(0.4,~0.6) < \pt < 4.05$\,\GeVc and $|\eta| < 0.8~(0.77,~0.85)$. To improve the quality of the tracks a minimum of 80 (70, 90) out of the 159 possible spatial points inside the TPC are required. The PID selections are applied by comparing the measured \dEdx and time-of-flight with the expected values for a proton candidate. The agreement is expressed in multiples ($n^\text{PID}_\sigma$) of the detector resolution $\sigma$. For protons with $\pt < 0.75$~\GeVc the $n^\text{PID}_\sigma$ is evaluated only based on the specific energy loss in the TPC, while for $\pt \geq 0.75$~\GeVc a combined TPC and TOF PID selection is applied $\left(n^\text{PID}_\sigma=\sqrt{n_{\sigma,\mathrm{TPC}}^2+n_{\sigma,\mathrm{TOF}}^2}\right)$. The $n^\text{PID}_\sigma$ of the accepted proton candidates is required to be lower than 3 (2.5, 3.5). To reject particles that are non-primary or come from pile-up collisions, the distance of closest approach (DCA) to the PV of the tracks is required to be less than 0.1~cm in the transverse plane and less than 0.2~cm along the beam axis. The purity of candidates is estimated using Monte Carlo (MC) simulations by taking the ratio of the number of reconstructed true protons produced by the generator and the number of all candidates identified as protons as a function of the reconstructed transverse momentum. The contributions of secondary protons stemming from weak decays of strange baryons and from interactions in the detector material are extracted using MC template fits to the measured distributions of the DCA to the PV~\cite{ALICE:Run1}. The average purity of the identified protons is $98.3\%$ and $86.6\%$ of them are primaries.

The \lmb candidates are reconstructed via the weak decay $\lmb\rightarrow \mathrm{p}\pi^-$ (the $\overline{\lmb}\rightarrow \overline{\mathrm{p}}\pi^+$ in case of $\overline{\Lambda}$ reconstruction). 
The secondary daughter tracks are selected with similar criteria as for the primary protons regarding $|\eta|$ and the number of hits in the TPC. However, a less strict PID requirement of $n^\text{PID}_\sigma<5(4)$ is used. In addition, the daughter tracks are required to have a DCA to the PV of at least 0.05~(0.06)~cm and the DCA between the daughter tracks at the secondary vertex must be smaller than 1.5~(1.2)~cm.
The cosine of the pointing angle (CPA) between the vector connecting the PV to the decay vertex and the 3-momentum of the \lmb candidate is required to be larger than 0.99~(0.995). To reject unphysical secondary vertices, reconstructed with tracks stemming from pile-up of pp collisions occurring in different bunch crossings, the decay tracks are required to possess a hit in the two innermost or the two outermost ITS layers or a matched TOF signal~\cite{ALICE:LL}. Finally, a selection on the candidate invariant mass (IM) is applied by requiring it to be in a $\pm\,4$ MeV/$c^2$ interval around the nominal $\Lambda$ mass~\cite{PDG}. The primary and secondary contributions to the yield of \lmb are extracted employing a similar method as for protons but using the CPA as an observable for the template fits. The \lmb hyperons produced in primary interactions contribute to about 58.5\% of their total yield. About 19.5\% originate from the electromagnetic decays of $\Sigma^0$.
The number of $\Sigma^0$ particles is related to their ratio to the \lmb hyperons, which is fixed to $1/3$ based on predictions from the isospin symmetry and a measurement of the corresponding production ratios~\cite{Albrecht:1986me}. Further, each of the weak decays of $\Xi^-$ and $\Xi^0$ contributes about 11 \% to the yield of \lmb hyperons. 
The purity of \lmb and \almb has been extracted by fitting the IM spectra of candidates as a function of the three-particle kinematic variable \Qthree which is defined in Eq.~\ref{eq:Q3general}. The fits have been performed in the IM range of 1090 to 1150 MeV/$c^2$ using a double Gaussian for the \lmb signal and a second-order polynomial for the background. The result has been averaged for $\Qthree<1$~\GeVc, leading to a combined purity of \lmb and \almb of 95.6\%.

The systematic uncertainties are evaluated by performing simultaneous variations of the selection criteria for protons and \lmb candidates as well as for the corresponding antiparticles. The variations are randomly combined in 44 sets in which at least one of the selection criteria is varied. Such procedure allows to account for the correlations between the systematic uncertainties. Each random set of variations is accepted for the evaluation of the systematic uncertainties only if the yield of the triplets is varied by less than 10\% with respect to the standard selection in the kinematic region $\Qthree<0.4$~\GeVc.

\subsection{Three-particle correlation function}
\label{3corr}
The observable of interest in femtoscopy is usually the two-particle momentum correlation function~\cite{Lisa:2005dd,Lednicky:1981su}, which is defined as the probability to simultaneously find two particles with momenta $\mathbf{p}_1$ and $\mathbf{p}_2$ divided by the product of the corresponding single particle probabilities
\begin{equation}
C(\mathbf{p}_1,\mathbf{p}_2)\equiv
\frac{P(\mathbf{p}_1,\mathbf{p}_2)}{P(\mathbf{p}_1) P(\mathbf{p}_2)}.
\label{eq:CFtheo}
\end{equation}
These probabilities are related to the inclusive Lorentz-invariant spectra $P(\mathbf{p}_1,\mathbf{p}_2) \propto E_1 E_2 \frac{\mathrm{d}^6N}{\mathrm{d}^3\mathbf{p}_1 \mathrm{d}^3\mathbf{p}_2}$ and $P(\mathbf{p}_{i}) \propto E_{i}\frac{\mathrm{d}^3N_{i}}{\mathrm{d}^3\mathbf{p}_{i}}$.
In the absence of a correlation signal, the value of $C(\mathbf{p}_1,\mathbf{p}_2)$ is constant and normalised to unity. A similar logic can be followed  to construct the three-particle correlation functions as

\begin{equation}
C(\mathbf{p}_1,\mathbf{p}_2,\mathbf{p}_3)\equiv
\frac{P(\mathbf{p}_1,\mathbf{p}_2,\mathbf{p}_3)}{P(\mathbf{p}_1) P(\mathbf{p}_2) P(\mathbf{p}_3)}.
\label{eq:CFtheo3}
\end{equation}

Following~\cite{koonin1977proton,pratt1990detailed}, Eq.~\ref{eq:CFtheo} can also be written as
\begin{equation}
C(\mathbf{k^*})= \int \, d^3 r^* S(r^*) |\psi(\mathbf{r}^*,\mathbf{k^*})|^2,
\label{eq:CFsourcewf}
\end{equation}
where $S(r^*)$ is the distribution of the relative distances of particle pairs in the pair rest frame (PRF, denoted by the $^*$) --- the so-called source function.
The properties of the source in pp collisions at $\sqrt{s}$ = 13 TeV have been evaluated in Ref.~\cite{Acharya:2020dfb}, including the effects of short-lived resonance decays which enlarge the effective source size. The wave function of the particle pair relative motion is denoted by $\psi(\mathbf{r}^*,\mathbf{k^*})$  where $\mathbf{k^*} = (\mathbf{p}^*_1-\mathbf{p}^*_2)/2$ is the relative momentum. The wave function encapsulates the details of the particle interaction and drives the shape of the correlation function. In case of the three-particle correlation function, the two-particle source function and the  wave function of the particle pair relative motion must be replaced by a three-particle source function and wave function. In this analysis, the measured three-particle correlation functions are not compared to theoretical predictions. The goal here is to extract the three-particle femtoscopic cumulants which provide experimental evidence of the existence, or the absence, of genuine three-particle correlations, as explained in Section~\ref{sec:cumulants}.

The three-particle correlation function can be written as
\begin{equation}
C(\mathbf{p}_1,\mathbf{p}_2,\mathbf{p}_3)=C(Q_3)=\mathcal{N}\frac{N_{\mathrm{s}}(Q_3)}{N_{\mathrm{m}}(Q_3)},
\label{eq:CFexp3}
\end{equation}
where $N_{\mathrm{s}}(Q_3)$ and $N_{\mathrm{m}}(Q_3)$ are the same-event and mixed-event distributions of three particle combinations (triplets) as a function of $Q_3$ and $\mathcal{N}$ is the normalisation parameter. The Lorentz-invariant variable $Q_3$ is defined in~\cite{Dhevan} as 
\begin{equation}
    Q_3 = \sqrt{-q_{12}^{2} - q_{23}^2 - q_{31}^2} \ ,
\label{eq:Q3general}
\end{equation}
where $q_{ij}$ is the norm of the four-vector~\cite{Lisa:2005dd}
\begin{equation}
  q_{ij}^{\mu}=\left(p_{i}-p_{j}\right)^{\mu}-\frac{\left(p_{i}-p_{j}\right) \cdot P_{ij}}{ P_{ij}^{2}} P_{ij}^{\mu},  \quad P_{ij} \equiv p_{i}+p_{j}, 
\end{equation}
which can be rewritten as
\begin{equation}
 q_{ij}^\mu = \frac{2\ m_j}{m_i + m_j}\ p_{i}^\mu - \frac{2\ m_i}{m_i + m_j}\  p_{j}^\mu \ .
 \label{eq:qij}
\end{equation}
 
Here $m_{i}$ and $m_{j}$ are the particle $i$ and $j$ masses, $p_{i}^\mu$ and $p_{j}^\mu$ are the particle four momenta, while $q_{ij}^{\mu}$ is the relative four-momentum of the pair $ij$. In the case of same mass particles, the term $\frac{\left(p_{i}-p_{j}\right) \cdot P_{ij}}{ P_{ij}^{2}} P_{ij}^{\mu}$ becomes 0. In the non-relativistic case $q_{ij}^2 = - 4 {k^*_{ij}}^2$, where $k^*_{ij}$ is the relative momentum of the $ij$ pair in the PRF.

The mixed-event sample is obtained using event-mixing techniques, in which the particle triplets of interest are generated by combining single particles stemming from three different events. To maintain the same acceptance effects as in the same event sample, the mixing procedure is conducted only for events with similar $z$ position of the primary vertex and multiplicity~\cite{ALICE:Run1}. 
Additionally, in order to correct for possible differences in terms of multiplicity distribution between same and mixed events, the yield of the latter is re-weighted in each multiplicity interval to have the same statistical weight as the distribution when particles are from the same event. To account for the two-track merging and splitting effects due to the finite two-track resolution in the same-event sample, a minimum value of the distance between two proton tracks (in case of \pLCF pairs, the proton from $\Lambda$ decay is considered along with the primary proton)
on the azimuthal-polar angles plane $\Delta \eta$--$\Delta \varphi$ is applied to both the same- and mixed-event samples. The default selection is $\Delta \eta^2 + \Delta \varphi^2 \geq0.017^2$ and a systematic variation of $+10$ \% for the value of the minimum distance is applied in the analysis. 
The normalisation parameter $\mathcal{N}$ is chosen
such that the mean value of the correlation function equals unity in a $Q_3$ region where the effects of FSIs are negligible. The interval $Q_3\in (1.0 - 1.2)$ GeV/$\textit{c}$ is chosen for all triplets. 
\subsection{Three-particle femtoscopic cumulants}\label{sec:cumulants}
The measurable three-particle correlation function $C(\mathbf{p}_1,\mathbf{p}_2,\mathbf{p}_3)$ include all interactions at work in the three-particle system: the  two-body interactions among all pairs within the selected triplet and the genuine three-body interaction. To access only the genuine three-body correlations, one can use cumulants.
Given random variables $X_i$, the cumulant for a triplet is defined by Kubo~\cite{Kubo} as
\begin{equation}
\begin{aligned}
\left\langle X_{1} X_{2} X_{3}\right\rangle_{\mathrm{c}} &=\left\langle X_{1} X_{2} X_{3}\right\rangle \\
&-\left\{\left\langle X_{1} X_{2}\right\rangle\left\langle X_{3}\right\rangle+\left\langle X_{2} X_{3}\right\rangle\left\langle X_{1}\right\rangle+\left\langle X_{3} X_{1}\right\rangle\left\langle X_{2}\right\rangle\right\} \\
&+2\left\langle X_{1}\right\rangle\left\langle X_{2}\right\rangle\left\langle X_{3}\right\rangle,
\end{aligned}
\label{eq:cumulantKubo}
\end{equation}
where $\left\langle X_{i}\right\rangle$ is the expectation value of the variable $X_i$ and $\left\langle X_{i} X_{j}\right\rangle$, $\left\langle X_{i} X_{j} X_{k}\right\rangle$ are the two- and three-variable joint moments. The three-particle correlation function, defined in Eq.~\ref{eq:CFexp3}, is the three-particle momentum distribution normalised to the mixed-event distribution. The cumulants method can be applied to the numerator which contains the correlated particles, and then the expression is normalised to the mixed-event distribution. The three-particle femtoscopic cumulant ${c}_{3}$ thus can be defined as 
\begin{equation}
\begin{aligned}
{c}_{3}\left(\textbf{p}_{1}, \textbf{p}_{2}, \textbf{p}_{3}\right) &=\left[ N_{3}\left(\textbf{p}_{1}, \textbf{p}_{2}, \textbf{p}_{3}\right) \right.\\
&-N_{2}\left(\textbf{p}_{1}, \textbf{p}_{2}\right) N_{1}\left(\textbf{p}_{3}\right)-N_{2}\left(\textbf{p}_{2}, \textbf{p}_{3}\right) N_{1}\left(\textbf{p}_{1}\right) -N_{2}\left(\textbf{p}_{3}, \textbf{p}_{1}\right) N_{1}\left(\textbf{p}_{2}\right) \\
&+ 2 N_{1}\left(\textbf{p}_{1}\right) N_{1}\left(\textbf{p}_{2}\right) N_{1}\left(\textbf{p}_{3}\right)\left.\right] / N_{1}\left(\textbf{p}_{1}\right) N_{1}\left(\textbf{p}_{2}\right) N_{1}\left(\textbf{p}_{3}\right),
\label{eq:cumulantFemto1}
\end{aligned}
\end{equation}
where $N_{3}\left(\textbf{p}_{1}, \textbf{p}_{2}, \textbf{p}_{3}\right)$ and $N_{2}\left(\textbf{p}_{i}, \textbf{p}_{j}\right)$ are the same-event three- and two-particle momentum distributions; $N_{1}\left(\textbf{p}_{i}\right)$ is the single-particle momentum distribution; the product terms $N_{2}\left(\textbf{p}_{i}, \textbf{p}_{j}\right)N_{1}\left(\textbf{p}_{k}\right)$ and $N_{1}\left(\textbf{p}_{i}\right)N_{1}\left(\textbf{p}_{j}\right)N_{1}\left(\textbf{p}_{k}\right)$ indicate the mixed event distributions.
Thus one can further rewrite the femtoscopic cumulant as
\begin{equation}
{c}_{3}\left(\textbf{p}_{1}, \textbf{p}_{2}, \textbf{p}_{3}\right) = C(\mathbf{p}_1,\mathbf{p}_2,\mathbf{p}_3) - C([\mathbf{p}_1,\mathbf{p}_2],\mathbf{p}_3) -C([\mathbf{p}_2,\mathbf{p}_3],\mathbf{p}_1)-C([\mathbf{p}_3,\mathbf{p}_1],\mathbf{p}_2)+2.
\label{eq:cumulantFemto}
\end{equation}

This method has been already successfully applied within the  ALICE Collaboration to study the possibility of coherent pion production by measuring three-pion femtoscopic cumulants in Ref.~\cite{Dhevan,PhysRevC.93.054908}. Theorem I from Ref.~\cite{Kubo} enunciates that the three-particle cumulant is zero if the variables $X_{i}, X_{j}, ...$ can be divided into two or more groups that are statistically independent.
 In case of femtoscopic cumulants, this translates into  $c_{3}\left(\textbf{p}_{1}, \textbf{p}_{2}, \textbf{p}_{3}\right)=0$ in the absence of genuine three-body correlations. Therefore, the measurements of non-vanishing values of $c_3$ can be used as an experimental confirmation of the existence of genuine three-body effects.\\
If genuine three-body correlations are not present in the particle triplet, the three-particle correlation function can be expressed using only lower order contributions as follows
\begin{equation}
    C^{\mathrm{two\text{-}body}} (\mathbf{p}_1,\mathbf{p}_2,\mathbf{p}_3)= C([\mathbf{p}_1,\mathbf{p}_2],\mathbf{p}_3) +C([\mathbf{p}_2,\mathbf{p}_3],\mathbf{p}_1)+C([\mathbf{p}_3,\mathbf{p}_1],\mathbf{p}_2)-2. 
    \label{eq:two-body}
\end{equation}
In Eq.~\ref{eq:two-body}, $C([\mathbf{p}_i,\mathbf{p}_j],\mathbf{p}_k)$ is built by combining particles $i$ and $j$ from the same event with particle $k$ from another event to obtain the numerator $\left(N_{2}\left(\textbf{p}_{i}, \textbf{p}_{j}\right) N_{1}\left(\textbf{p}_{k}\right)\right)$ of the correlation function while the denominator $\left(N_{1}\left(\textbf{p}_{1}\right) N_{1}\left(\textbf{p}_{2}\right) N_{1}\left(\textbf{p}_{3}\right)\right)$ is estimated using three particles from three different events as described in Section~\ref{3corr}.

\subsection{Projector method}
\label{sec:projector}
An alternative method to isolate the genuine three-body contribution to the measured three-particle correlation functions is the projector method~\cite{DelGrande:2021mju}.  This method makes use of the subtraction rule provided by the Kubo's cumulant decomposition (Eq.~\ref{eq:cumulantFemto}) but, instead of evaluating them with the data-driven approach based on event mixing described above, it calculates $C([\mathbf{p}_i,\mathbf{p}_j],\mathbf{p}_k)$ using the measured or the calculated two-particle correlation function and the projection of the third non-interacting (spectator) particle. The method is described in Ref.~\cite{DelGrande:2021mju}. Given the three-particle correlation function, $C(Q_3)$, and the two-body correlation functions, $C(k^{*}_{ij})$, the projector method provides a kinematic transformation from the relative momentum $k^{*}_{ij}$ of the interacting pairs $ij$ to the $Q_3$ of the three-body system $(i-j)-k$ under study. 
The transformation is given by the following integral in the momentum space
\begin{equation}
       C_{ij}(Q_3) = \int C(k^{*}_{ij}) \ W_{ij}(k^{*}_{ij},Q_3) \ dk^{*}_{ij},
       \label{eq:C3ij}
\end{equation}
where the indices $ij$ denote the interacting pair and the projector function $W_{ij}$ is equal to~\cite{DelGrande:2021mju} 
\begin{equation}
 W_{ij}(k^{*}_{ij},Q_3) = \frac{16 (\alpha \gamma - \beta^2)^{3/2} {k^{*}_{ij}}^2}{\pi Q_3^4 \gamma^2} \sqrt{\gamma Q_3^2 - (\alpha \gamma - \beta^2) {k^{*}_{ij}}^2}.
    \label{projector}
\end{equation}
The constants $\alpha$, $\beta$ and $\gamma$ depend on the particle masses\footnote{ $\alpha = \frac{4\ m_k^2}{(m_i+m_k)^2}+\frac{4\ m_k^2}{(m_j+m_k)^2}+ 4 \ ;$ $\beta = \frac{4\ m_k (m_i+m_j+m_k)}{m_i+m_j}
   \left[ \frac{m_j}{(m_j+m_k)^2}-\frac{m_i}{(m_i+m_k)^2}\right] \ ;$ $\gamma  = \frac{4\ (m_i+m_j+m_k)^2}{(m_i + m_j)^2}
   \left[\frac{ m_i^2}{(m_i + m_k )^2}+\frac{m_j^2}{( m_j + m_k)^2}\right] \ .$}.
\begin{figure}[ht]
\centering
\includegraphics[width=0.32\textwidth]{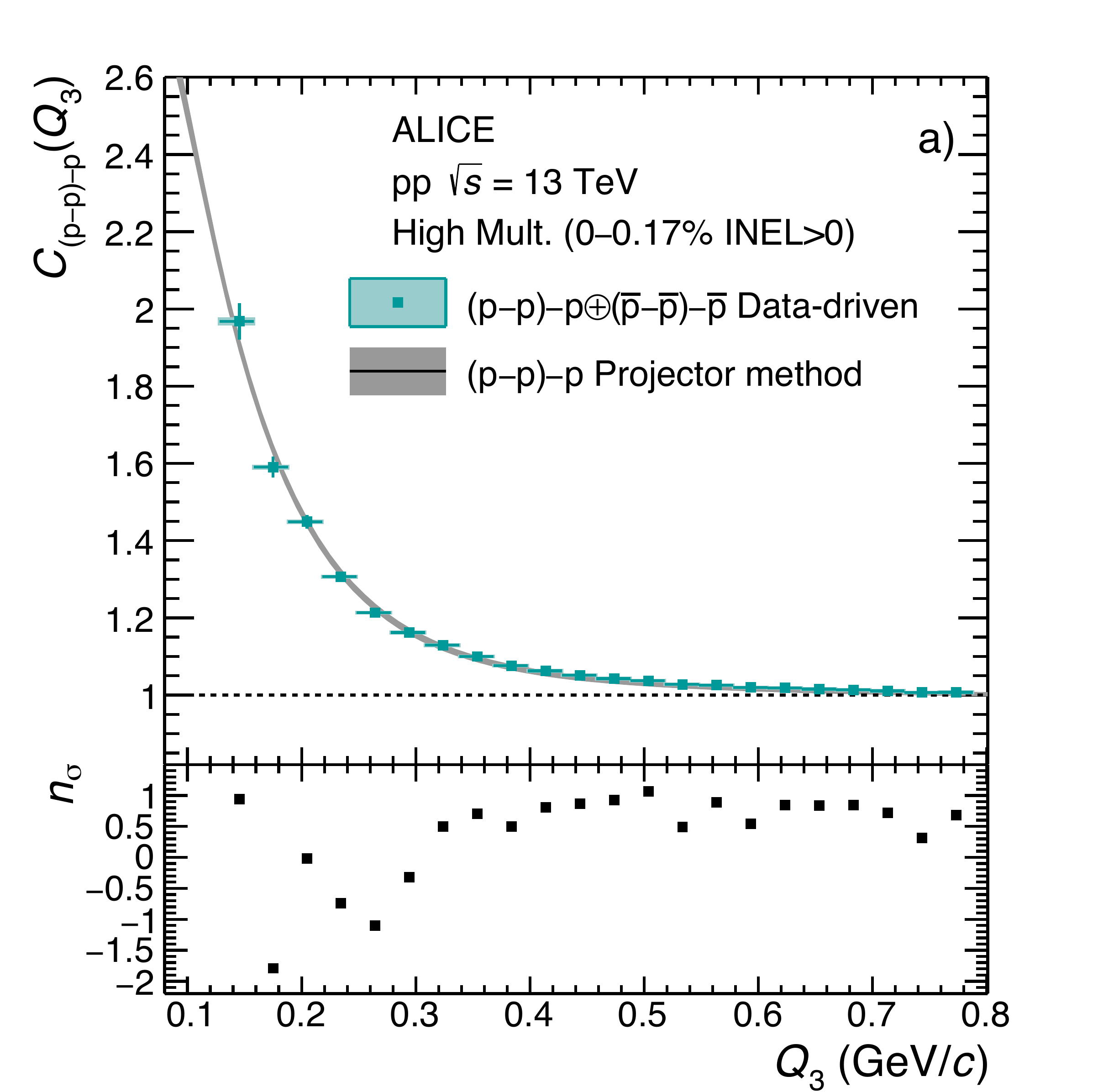}
\includegraphics[width=0.32\textwidth]{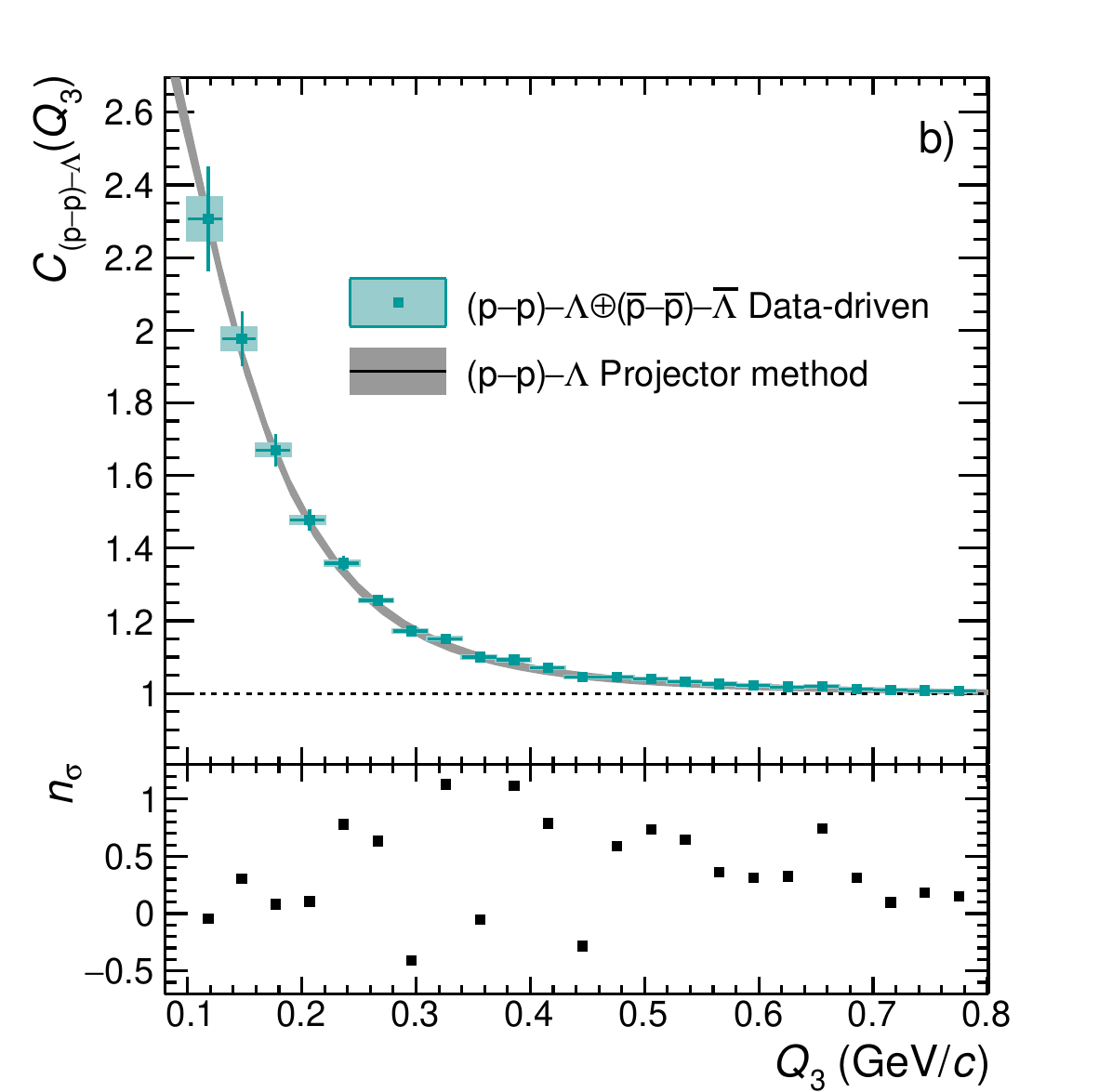}
\includegraphics[width=0.32\textwidth]{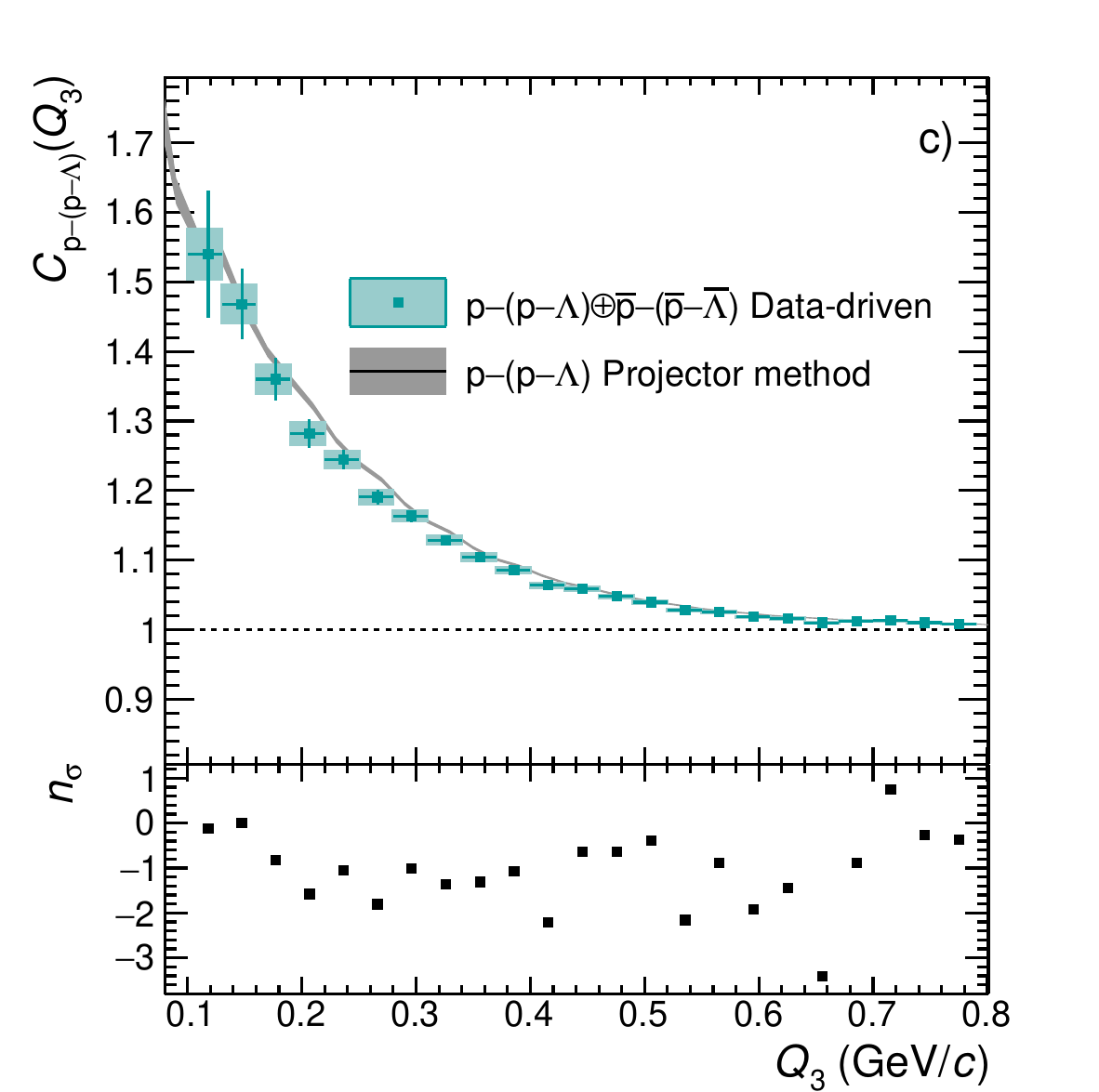}
\caption{The upper panels show the comparison of the two-particle correlations projected on three-particle phase space obtained using the data-driven approach based on event mixing (green points) and the projector method (grey band). The resulting correlation functions are shown for \ppspCF (panel a), \ppsLCF (panel b) and \ppLsCF  (panel c) cases. The error bars and the boxes represent the statistical  and systematic uncertainties, respectively. The grey band includes systematic and statistical uncertainties summed in quadrature. The lower panels show the deviations between the data-driven approach and the projector method, expressed in terms of $n_\sigma$.}
\label{fig:2bodyseparate}
\end{figure}
The integral in Eq.~\ref{eq:C3ij} can be evaluated using the measured p--p and p--$\Lambda$ correlation functions from Refs.~\cite{ALICE:pSig0,Acharya:2020dfb,ALICE:2021njx}. The resulting correlation functions
are compared to the ones obtained by employing the data-driven method (Eq.~\ref{eq:two-body}) and shown in Fig.~\ref{fig:2bodyseparate}. Panel a) shows the \ppspCF correlation function, the green points are obtained using the data-driven approach and the grey band is obtained with the projector method. The statistical and systematic uncertainties are shown separately for the data driven method, while the width of the grey band represents the sum in quadrature of the statistical and systematic uncertainties for the projector method. The statistical uncertainties of all the measured correlation functions have been estimated using a bootstrap~\cite{ALICE:2021njx} method by sampling same- and mixed-event counts from Poisson distributions. The statistical uncertainties shown correspond to the central 68\% confidence interval and are consistent with the uncertainties obtained employing the standard error propagation method. The systematic uncertainties are estimated by varying the selection criteria of the particle candidates as described in Section~\ref{sec:eventselection}. Panels b) and c) show the same comparison for the \ppsLCF and the \ppLsCF correlation functions. The number of events used for mixing to obtain \ppspCF and \ppsLCF correlation functions is 30. The numerator of \ppLsCF correlation function requires p--$\Lambda$ pairs in same event sample, which are less abundant than p--p pairs. Thus to obtain good statistical precision, the number of events used for mixing (to account for the third uncorrelated particle) must be increased to 100 in case of \ppLsCF triplets.

The results from the data-driven and the projector method are in good agreement between each other. The number of deviations $n_{\sigma}$ in each bin are shown in the bottom panels of Fig.~\ref{fig:2bodyseparate}, where $\sigma$ is the combined statistical and systematic uncertainty for both the experimental data and the projector. The agreement in the region $Q_3 < 0.8$ GeV/$c$ has been evaluated by performing a $\chi^2$ test. The $\chi^2$ is calculated combining the $n_{\sigma}$ values of each bin. Finally, the p-value from the $\chi^2$-distribution is computed and the global  $n_{\sigma}$ values are extracted. The latter amount to 0.167, 0.0006 and 2.75 for \ppspCF, \ppsLCF and \ppLsCF, respectively.
The data-driven method requires the usage of the third particle in the triplet from the mixed-event data sample and consequently the statistical uncertainty  depends on the number of events used for mixing, while the projector method does not have this limitation.  Thus, the latter significantly reduces the total uncertainty in the evaluation of the two-particle correlation effect on the three-particle correlation functions. For this reason, the projector method is used to calculate the three-particle cumulants for the \pppCF and \ppLCF triplets.

\begin{figure}[ht]
\centering
\includegraphics[width=0.49\textwidth]{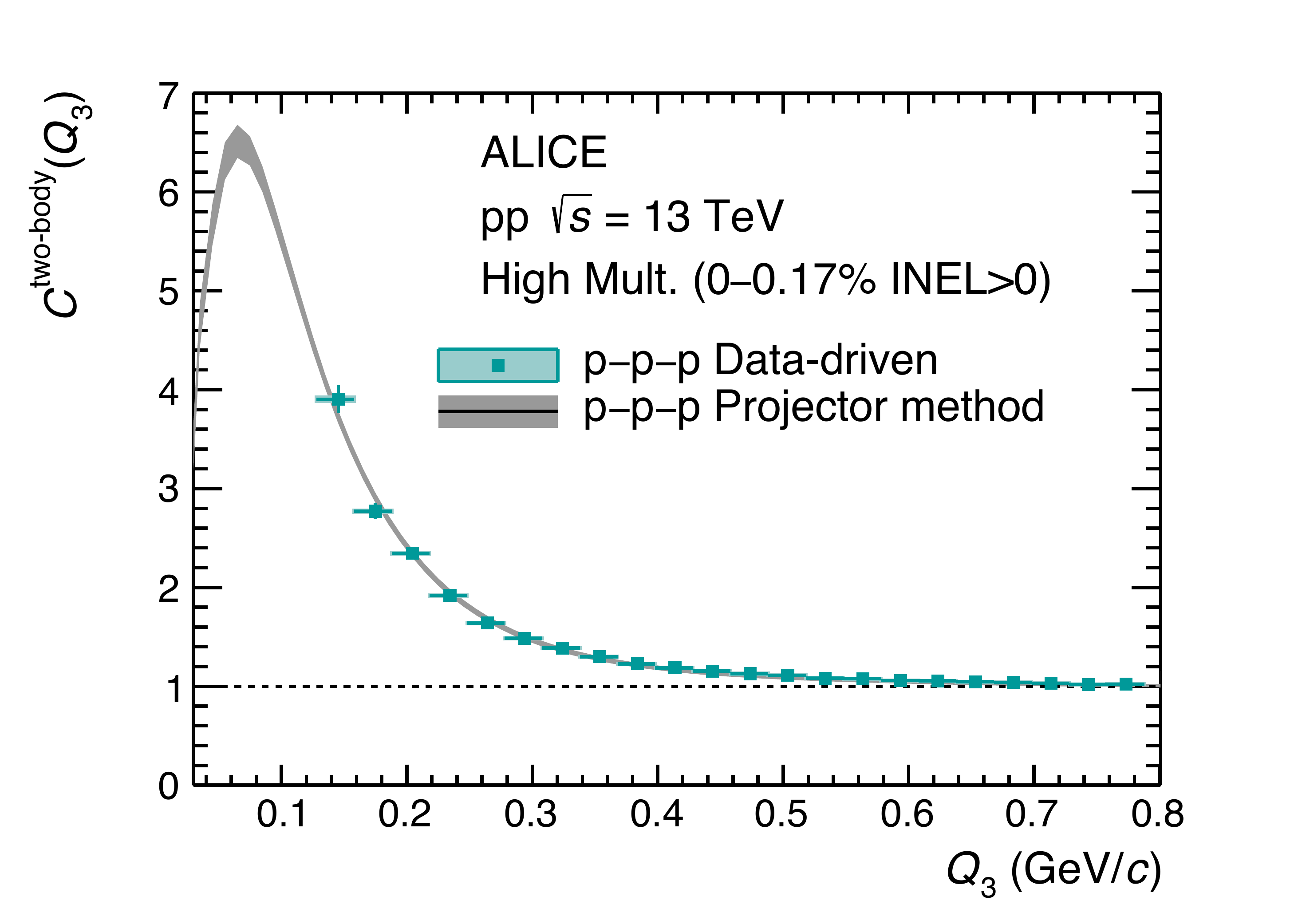}
\includegraphics[width=0.49\textwidth]{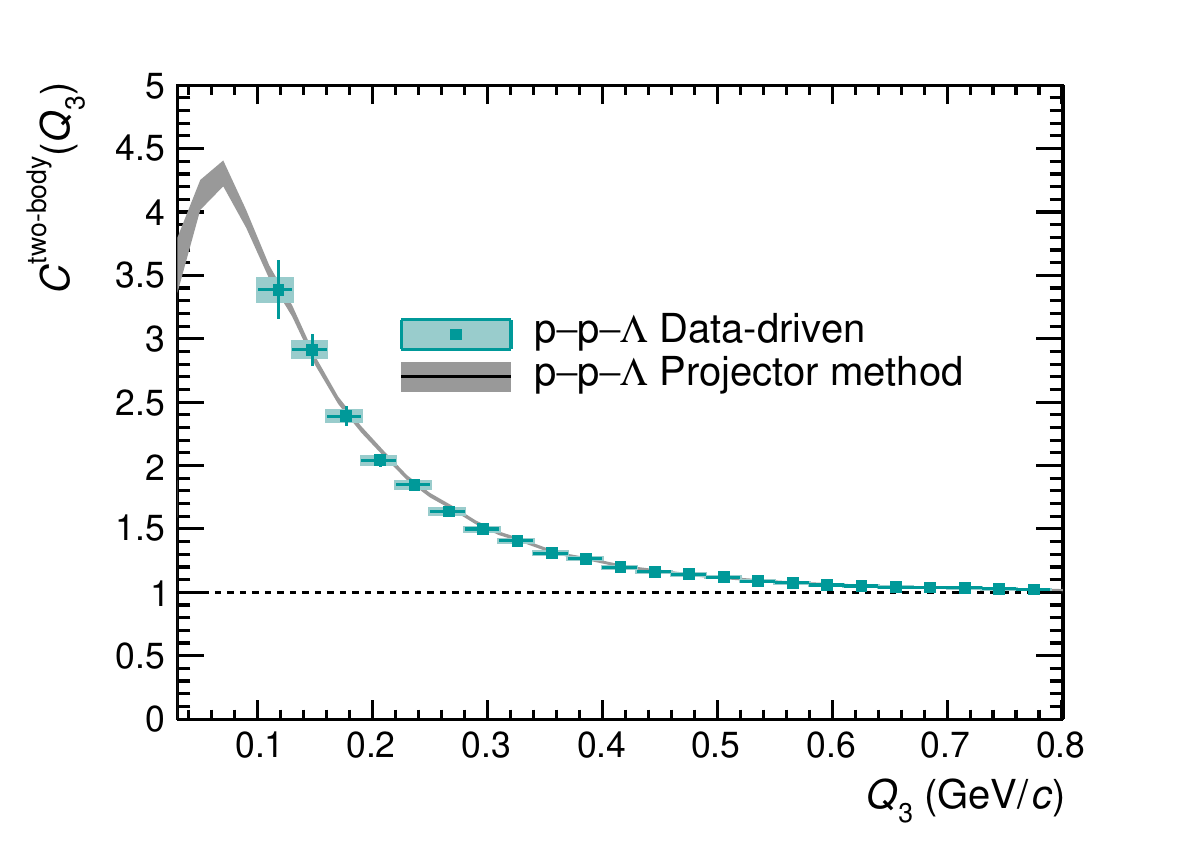}
\caption{Comparison of the total two-particle contribution to the three-particle correlation functions obtained using the data-driven approach (green points) and the projector method (grey band). The resulting correlation functions are shown for \pppCF (left panel) and \ppLCF (right panel). The error bars and the boxes represent the statistical  and systematic uncertainties, respectively. The grey band includes systematic and statistical uncertainties summed in quadrature.}
\label{fig:lower_order}
\end{figure}

The total two-particle contribution to the three-particle correlation function is obtained by substituting all terms on the right-hand side of Eq.~\ref{eq:two-body} with the corresponding kinematic transformation, i.e.
\begin{equation}
\label{eq:projected3body}
        C^{\mathrm{two\text{-}body}} (Q_3) = C_{12} (Q_3) + C_{23} (Q_3) + C_{31} (Q_3) - 2  \ ,
\end{equation}
where the indices refer to the label of the correlated pairs.
In the case of \pppCF we have
\begin{equation}
    C_\mathrm{p-p-p}^\mathrm{two\text{-}body} (Q_3) =  3 \ C_\mathrm{(p-p)-p} (Q_3) - 2 \ ,
    \label{eq:formPPP}
\end{equation}
and in the case of \ppLCF we have
\begin{equation}
    C_\mathrm{p-p-\Lambda}^\mathrm{two\text{-}body} (Q_3) =  C_\mathrm{(p-p)-\Lambda} (Q_3) + 2 \ C_\mathrm{p-(p-\Lambda)} (Q_3) - 2 \ .
    \label{eq:formPPL}
\end{equation}
The resulting total lower-order contributions to the three-particle correlation functions (Eqs.~\ref{eq:formPPP} and~\ref{eq:formPPL}) are shown in Fig.~\ref{fig:lower_order}. The agreement between the data-driven approach and the projector method predictions translate into $n_{\sigma}$ = 0.167 and $n_{\sigma}$ = 0.0014 for the \pppCF and \ppLCF lower-order contributions, respectively.

\subsection{Decomposition of the three-particle cumulants}
\label{sec:decomposition}
The experimental determination of the correlation function is mainly distorted by two distinct impurities in the candidate sample: misidentified particles and feed-down particles originating from weakly decaying particles. This introduces additional contributions to the correlation function of interest. These contributions are either assumed to be flat or, when the interaction is known, they are explicitly considered as discussed in Ref.~\cite{ALICE:Run1}. The contributions to the correlation function stemming from decaying particles or impurities of the sample are weighted with the so-called $\lambda$ parameters. By adopting this technique the residual correlations can be included in the final description of the experimental correlation function of two particles as 
\begin{equation}\label{eq:Ck_lambda}
C(k^*) = 1 + \lambda_{\mathrm{00}} ( C_{\mathrm{00}}(k^*)-1 )
+\sum_{ij\neq 00}\lambda_{ij}(C_{ij}(k^*) -1),
\end{equation}
where the $ij\neq 00$ denote all possible impurity and feed-down contributions and the $ij=00$ is the correctly identified primary particle contribution. These $\lambda$ parameters are obtained employing single particle properties such as the purity and feed-down probability. The underlying mathematical formalism is outlined in Ref.~\cite{ALICE:Run1}. This mechanism has been extended to the three-particle case and the genuine three-particle cumulants can be obtained by subtracting the impurity and feed-down contributions from the measured cumulants. The full mathematical derivation is presented in Appendix~\ref{sec:AppendixB}.
The final expression of the genuine three-particle cumulants is 
\begin{equation}
\begin{aligned}
c(X_0 Y_0 Z_0) &= \frac{1}{\lambda_{X_0 Y_0 Z_0}(X YZ)} \left(c(XYZ) - \sum_{i, j,k\neq (X_0 Y_0 Z_0)} \lambda_{i, j,k}(X Y Z) c(X_i Y_j Z_k) \right) \ ,
\end{aligned}
\label{eq:derivecfBEAUTY}
\end{equation}
where $X$, $Y$ and $Z$ represent three generic particle species, the index $0$ refers to correctly identified primary particle and the indexes $i,j,k$ refer to misidentified or to secondary particles of a generic particle species. As shown in Appendix~\ref{sec:AppendixB}, the specific weights $\lambda$ depend on the purity and feed-down fraction of the single particles and are found to be equal to $\lambda_{X_0 Y_0 Z_0}(\mathrm{ppp})=\, 0.618$ and  $\lambda_{X_0 Y_0 Z_0}(\mathrm{pp\Lambda})=\,0.405$ for the \pppCF and \ppLCF cumulants, respectively. 
Only 60\% (40\%) of the \pppCF (\ppLCF) triplets correspond to correctly identified primary particles. 
 
 In the following, the results for the \pppCF cumulants will be corrected according to the evaluated $\lambda$ parameters assuming that all the three-particle contributions stemming from feed-down and impurities are flat in the momentum space. This assumption is supported by the observation that the measured \ppLCF cumulants are consistent with zero within uncertainties (see Fig.~\ref{fig:cumulants} and the discussion in Section~\ref{sec:results}). The correction is not applied to the \ppLCF cumulants because the shape of the feed-down contribution is not known and also because the statistical uncertainties are too large to provide any sensitivity to the three particle correlations.

\section{Results}
\label{sec:results}
\begin{figure}[ht]
\centering
\includegraphics[width=0.48\textwidth]{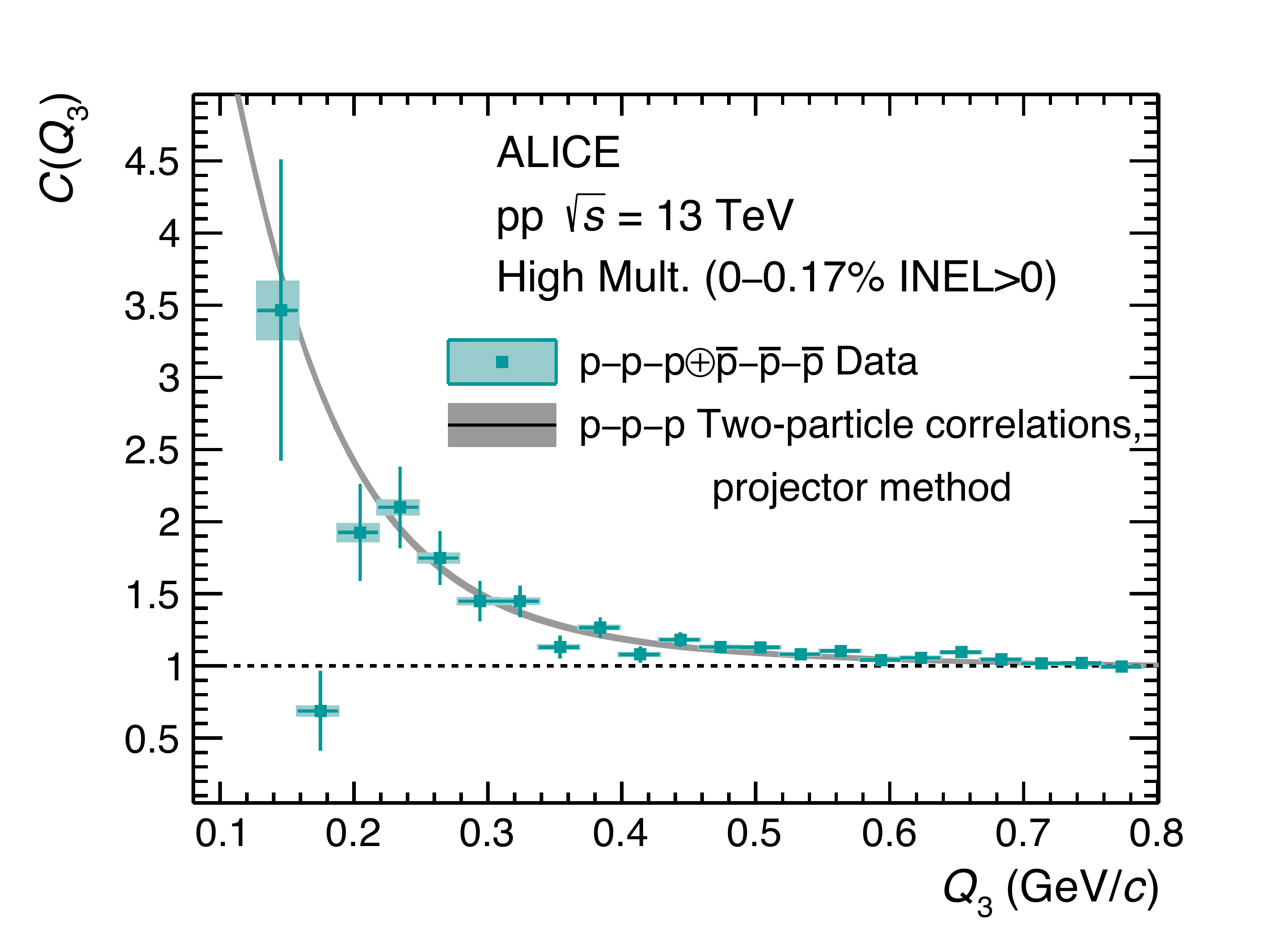}
\includegraphics[width=0.48\textwidth]{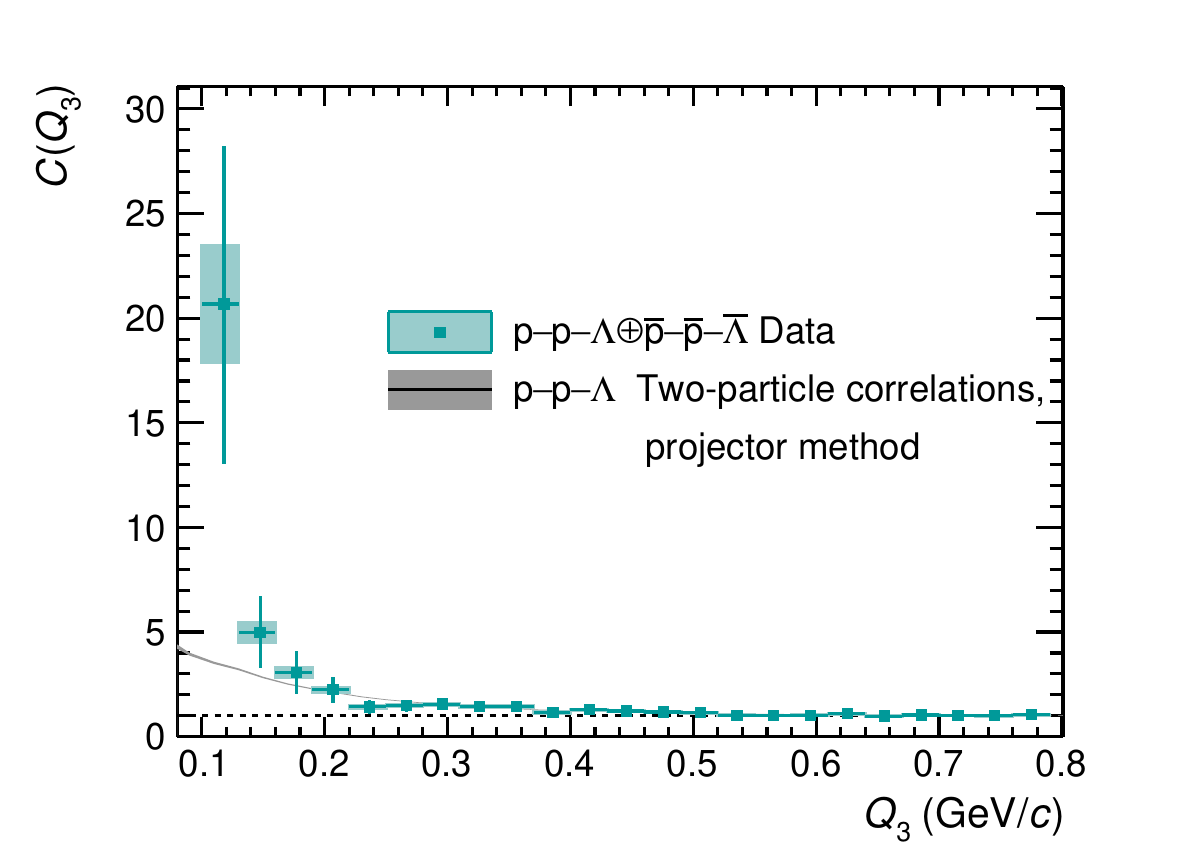}
\caption{Measured \pppCF (left panel) and \ppLCF (right panel) three-particle correlation functions. The green points show the experimental results, the error bars and the boxes represent the statistical and systematic uncertainties, respectively. The grey bands represent the expectations for the lower-order two-particle correlations obtained using the projector method and the band width is obtained including systematic and statistical uncertainties summed in quadrature.}
\label{fig:nMM}
\end{figure}

 The measured three-particle correlation functions for \pppCF and \ppLCF triplets are shown in Fig.~\ref{fig:nMM} on the left and right panels, respectively.  The number of events used for mixing for both cases is 30. The total number of same event triplets that are present at the range $Q_{3}<$ 0.8 GeV/$c$ are 17840 for \pppCF, 10980 for \apapapCF,  9191 for \ppLCF and 5886 for \apapaLCF.  The green symbols represent the data points with their statistical and systematic uncertainties, while the grey bands correspond to the lower-order two-body interaction contributions obtained using the projector method 
already shown in Fig.~\ref{fig:lower_order}. The non-femtoscopic contributions to the measured correlation functions, evaluated using Monte Carlo simulations, are found to be negligibly small (see Appendix~\ref{app:MC} for a detailed discussion).

\begin{figure}[ht]
\centering
\includegraphics[width=0.48\textwidth]{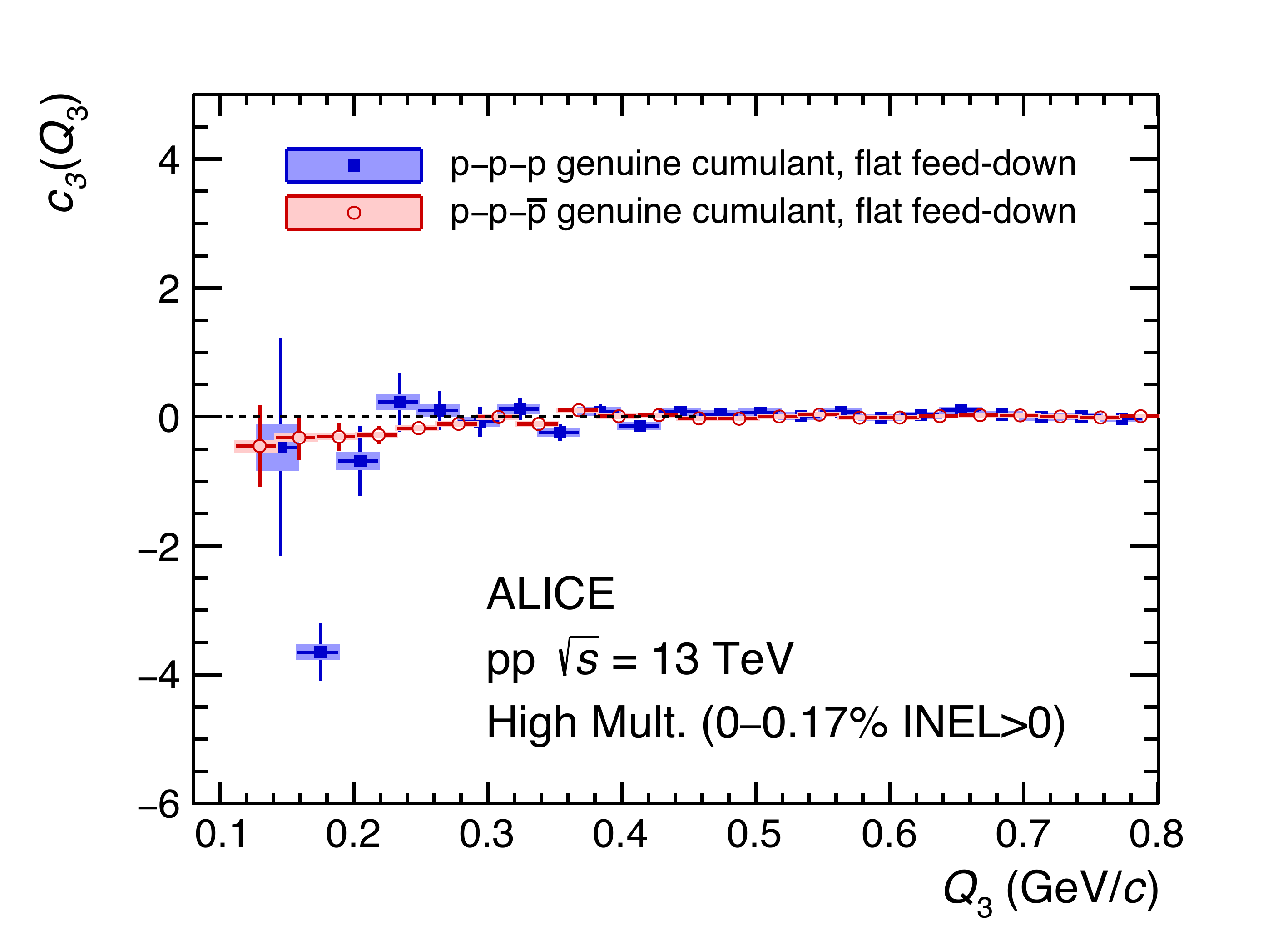}
\includegraphics[width=0.48\textwidth]{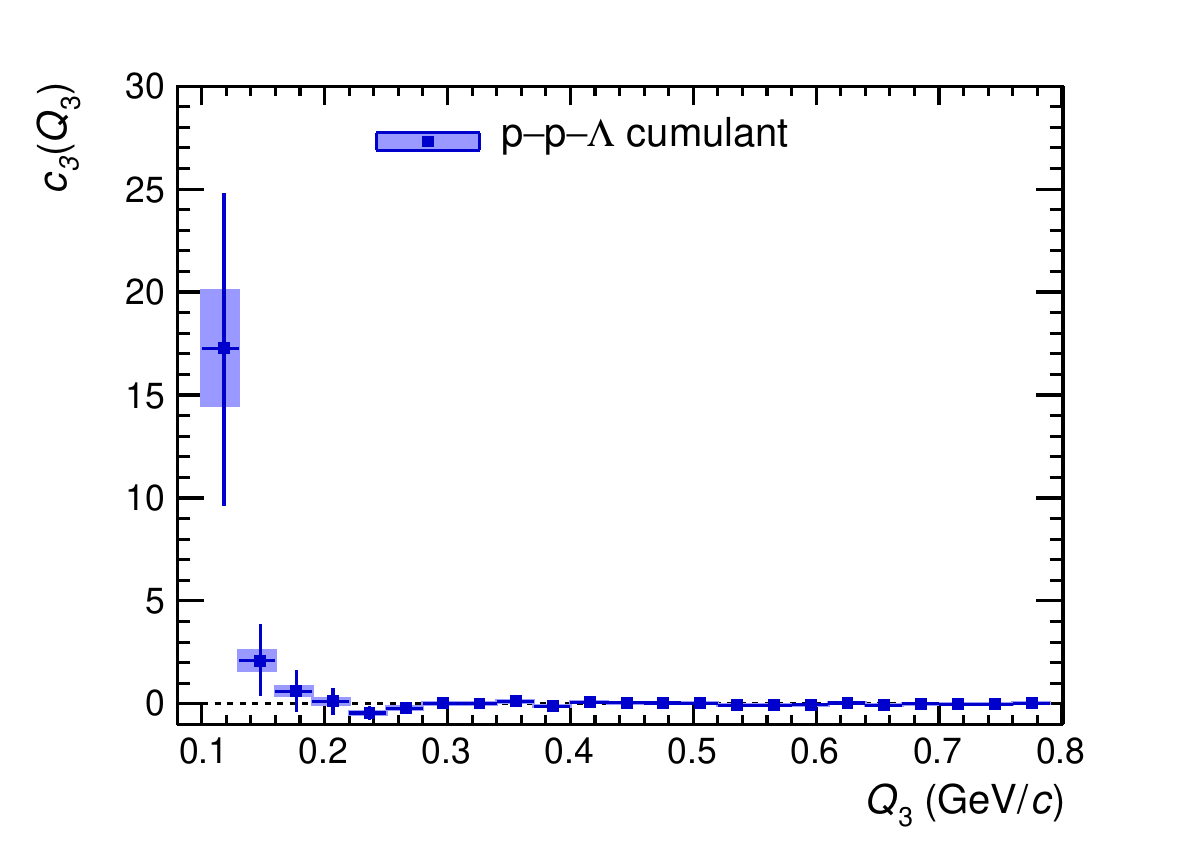}
\caption{Three-particle cumulants for \pppCF (left panel, blue square symbols) and \ppLCF (right panel) triplets obtained by subtracting the lower-order contributions from the measured three-particle correlation functions shown in Fig.~\ref{fig:nMM}. The \pppCF cumulant in the left panel is further corrected for the feed-down contributions from decaying particles and represents thus, the cumulant for the correctly identified primary protons (see Section~\ref{sec:decomposition} for details). The dashed lines correspond to the assumption that there are no genuine three-body correlations ${c}_3 (Q_3) = 0$. The red open circles in left panel represent the cumulant for \pppbarCF triplets (for more details see the main text).}
\label{fig:cumulants}
\end{figure}

In the low $Q_3$ region, the measured correlation functions deviate from the projected lower order contributions obtained using only two-particle correlations.
The genuine three-body effects 
are then isolated by evaluating the cumulants
\begin{equation}
    c_3 (Q_3) = C(Q_3) - C^\mathrm{two\text{-}body}(Q_3) \ .
\end{equation}
The lower-order contribution $C^\mathrm{two\text{-}body}(Q_3)$ obtained with the projector method is used. The results for  \pppCF and \ppLCF triplets are shown in Fig.~\ref{fig:cumulants} on the left and right panels, respectively.
The \pppCF cumulant, already corrected for the feed-down contributions, is negative for $0.16 < Q_3 < 0.22$ GeV/{\it c}, while the large statistical uncertainty in the lowest $Q_3$ interval prevents a conclusion on the sign for $Q_3 < 0.16$ GeV/{\it c}. The agreement between the measured cumulant and the assumption that there are no genuine three-body effects is evaluated using the $\chi^2$ test in the region $Q_3 < 0.4$ GeV/{\it c}, where the two-body interactions are prominent. There is no theoretical or experimental knowledge on the exact $Q_3$ range where three-body effects become relevant, however they are expected to contribute at lower or same $Q_3$ values as the two-body interactions. For this reason, the region of two-body forces was chosen. The obtained p-value corresponds to 6.7 standard
deviations. If the cumulant is obtained using data-driven method to estimate lower order contributions, it results in 6.0 standard deviations.
This result hints to the presence of an effect beyond the two-body interactions in the \pppCF system that could be either due to Pauli blocking (Fermi-Dirac quantum statistics) at the three particle level~\cite{Niemann:2012gd} or to the contribution of a three-body nuclear repulsive interactions. Long-range Coulomb interactions may also lead to significant contributions~\cite{alt1993asymptotic}. More quantitative conclusions on the interpretation of the non-zero cumulant require more sophisticated calculations for the three-body system. The present analysis demonstrates the experimental accessibility of the three-baryon cumulant in the data sample of pp collisions at $\sqrt{s}=13$ TeV recorded by ALICE.
In addition to the \pppCF system, the mixed-charge \pppbarCF case has also been studied (see Appendix~\ref{app:MixedChargeCumulant} for details). In the case of \pppbarCF, the two-body \ppbarCF interaction contains both elastic and inelastic components, previously measured in an independent analysis~\cite{ALICE:2021cyj}; the \ppCF interaction is well known from scattering data~\cite{Arndt:2007qn,NavarroPerez:2013usk} and verified by correlation measurements~\cite{ALICE:Run1,ALICE:pSig0,Acharya:2020dfb}; and the genuine three-body strong interaction for the \pppbarCF triplet should be negligible. The extracted cumulant for \pppbarCF is shown in the left panel of Fig.~\ref{fig:cumulants} by the red open circles. Since the number of the mixed-charge triplets is a factor four higher than the one of the same-charge triplets, it is possible to extend the measurement of the three-particle correlation to lower $Q_3$ values. The correction for the feed-down contributions has been applied and the statistical and systematic uncertainties are shown. The \pppbarCF cumulant evaluated using the projector method (data-driven approach) agrees with the assumption of only two-body correlations present in the system within 2.1 (2.2) standard deviations in the region 
$Q_3 < 0.4$ GeV/{\it c} and within 0.9 (0.9) standard deviations in the region 
$Q_3 < 0.2$ GeV/{\it c}, suggesting that genuine three-body effects are not statistically significant. This result as well demonstrates that the measured \pppCF cumulant deviation from zero is not due to detector effects.

In the case of \ppLCF, a positive cumulant is measured at $Q_3 < 0.16$ GeV/{\it c}. 
The p-value obtained from the $\chi^2$ test in the region $Q_3 < 0.4$ GeV/{\it c} corresponds to a deviation of 0.8 $\sigma$ from the assumption that no genuine three-body correlations are present. Using data-driven method to estimate lower order contributions results in 0.8 $\sigma$ as well.
A similar value is found by repeating the significance test with the Fisher method, meaning that the measured cumulant is compatible with the assumption of no genuine three-body effects within the uncertainties. The current measurement does not allow to draw any firm conclusion yet on the three-body interaction in the \ppLCF system, but since in this case only two of the particles are identical and charged, a non zero cumulant can be directly linked to the presence of a strong three-body interaction.
It is estimated that employing a three-baryon event filtering during the upcoming Run 3 data taking should increase the number of triplets by a factor up to 500 for the target integrated luminosity of 200 nb$^{-1}$ at $\sqrt{s}=13.6$ TeV~\cite{ALICE-pp-publicnote}. This opens up the possibility of measuring precisely the three-body correlations for both the \pppCF and \ppLCF systems.

\section{Conclusions}
\label{sec:conclusions}
In this article, the first femtoscopic study of the \pppCF and \ppLCF systems measured in high-multiplicity pp collisions at $\sqrt{s} = 13$ TeV with the ALICE detector has been presented. In the chosen colliding system, hadrons are emitted at average relative distances of about 1 fm providing a unique environment to test three-body interactions at scales shorter than inter-particle ones in nuclei. 
The data collected during the LHC Run 2 enabled the measurement of the \pppCF and \ppLCF correlation functions in the low $Q_3$ region down to 0.1 GeV/$c$, 
giving access to the region where the effects of the hadronic two- and three-body interactions are more pronounced. 
The genuine three-particle correlations have been isolated using the Kubo's cumulant expansion method.
The lower-order two-body contributions have been estimated employing both a data-driven event mixing technique and a newly developed projector method. The two approaches have been compared and found to be in good agreement between each other, providing the first validation of the projector method using the data. The extracted \pppCF and \ppLCF cumulants deviate from zero in the low $Q_3$ region. In the case of \pppCF, a negative three-particle cumulant is measured. The p-value extracted from the $\chi^2$ test corresponds to a deviation of 6.7$\sigma$ from the assumption of only two-body correlations present in the system for $Q_3 < 0.4$ GeV/$c$. 
The obtained result provides an experimental hint to the presence of an effect beyond pairwise interactions in the \pppCF system. The observed deviation could be due to genuine three-body effects arising from: Pauli blocking, short-range strong interactions, or long-range Coulomb interactions.
Refined three-body system calculations are required to give a solid interpretation of the measurement.
The mixed-charge \pppbarCF cumulant has also been measured as a benchmark and the result is consistent with the assumption that only two-body correlations are present in the system showing that the effect observed for the \pppCF system is a genuine one.
In the case of \ppLCF, a positive cumulant is observed at low $Q_3$.
The deviation from zero at $Q_3 < 0.4$ GeV/$c$ is 0.8 $\sigma$, which suggests no significant deviation from the assumption that only two-body correlations are present in the system within the current uncertainties.
For this system, where one particle is uncharged and only two particles are identical, genuine three-particle correlations can be directly linked to the three-body strong interaction. 
The upcoming LHC Run 3 data taking will provide significantly larger samples of measured triplets, allowing more quantitative conclusions to be drawn for many-body dynamics.

The analysis presented in this article represents a first important step towards the direct measurement of the three-body interaction among baryons, demonstrating that genuine three-particle effects can be studied using three-particle correlation functions as experimental observables.

%%%%%%%%%%%%%%%%%%%%%%%%%%%%%%%%%%%%%%%%%%%%%%%%%%%%%%%%%%%%%
\newenvironment{acknowledgement}{\relax}{\relax}
\begin{acknowledgement}
\section*{Acknowledgements}
The ALICE Collaboration is grateful to Prof. Alejandro Kievsky for the
fruitful discussions on the theoretical aspects of three-body systems.
%TC:ignore
% add specific acknowledgements here 
% ...but please don't remove the line below: funding agencies
% will be acknowledged with a custom tex file handled by EB chairs after Collab Round 2
% Version: 2022-05-20

The ALICE Collaboration would like to thank all its engineers and technicians for their invaluable contributions to the construction of the experiment and the CERN accelerator teams for the outstanding performance of the LHC complex.
The ALICE Collaboration gratefully acknowledges the resources and support provided by all Grid centres and the Worldwide LHC Computing Grid (WLCG) collaboration.
The ALICE Collaboration acknowledges the following funding agencies for their support in building and running the ALICE detector:
A. I. Alikhanyan National Science Laboratory (Yerevan Physics Institute) Foundation (ANSL), State Committee of Science and World Federation of Scientists (WFS), Armenia;
Austrian Academy of Sciences, Austrian Science Fund (FWF): [M 2467-N36] and Nationalstiftung f\"{u}r Forschung, Technologie und Entwicklung, Austria;
Ministry of Communications and High Technologies, National Nuclear Research Center, Azerbaijan;
Conselho Nacional de Desenvolvimento Cient\'{\i}fico e Tecnol\'{o}gico (CNPq), Financiadora de Estudos e Projetos (Finep), Funda\c{c}\~{a}o de Amparo \`{a} Pesquisa do Estado de S\~{a}o Paulo (FAPESP) and Universidade Federal do Rio Grande do Sul (UFRGS), Brazil;
Bulgarian Ministry of Education and Science, within the National Roadmap for Research Infrastructures 2020-2027 (object CERN), Bulgaria;
Ministry of Education of China (MOEC) , Ministry of Science \& Technology of China (MSTC) and National Natural Science Foundation of China (NSFC), China;
Ministry of Science and Education and Croatian Science Foundation, Croatia;
Centro de Aplicaciones Tecnol\'{o}gicas y Desarrollo Nuclear (CEADEN), Cubaenerg\'{\i}a, Cuba;
Ministry of Education, Youth and Sports of the Czech Republic, Czech Republic;
The Danish Council for Independent Research | Natural Sciences, the VILLUM FONDEN and Danish National Research Foundation (DNRF), Denmark;
Helsinki Institute of Physics (HIP), Finland;
Commissariat \`{a} l'Energie Atomique (CEA) and Institut National de Physique Nucl\'{e}aire et de Physique des Particules (IN2P3) and Centre National de la Recherche Scientifique (CNRS), France;
Bundesministerium f\"{u}r Bildung und Forschung (BMBF) and GSI Helmholtzzentrum f\"{u}r Schwerionenforschung GmbH, Germany;
General Secretariat for Research and Technology, Ministry of Education, Research and Religions, Greece;
National Research, Development and Innovation Office, Hungary;
Department of Atomic Energy Government of India (DAE), Department of Science and Technology, Government of India (DST), University Grants Commission, Government of India (UGC) and Council of Scientific and Industrial Research (CSIR), India;
National Research and Innovation Agency - BRIN, Indonesia;
Istituto Nazionale di Fisica Nucleare (INFN), Italy;
Japanese Ministry of Education, Culture, Sports, Science and Technology (MEXT) and Japan Society for the Promotion of Science (JSPS) KAKENHI, Japan;
Consejo Nacional de Ciencia (CONACYT) y Tecnolog\'{i}a, through Fondo de Cooperaci\'{o}n Internacional en Ciencia y Tecnolog\'{i}a (FONCICYT) and Direcci\'{o}n General de Asuntos del Personal Academico (DGAPA), Mexico;
Nederlandse Organisatie voor Wetenschappelijk Onderzoek (NWO), Netherlands;
The Research Council of Norway, Norway;
Commission on Science and Technology for Sustainable Development in the South (COMSATS), Pakistan;
Pontificia Universidad Cat\'{o}lica del Per\'{u}, Peru;
Ministry of Education and Science, National Science Centre and WUT ID-UB, Poland;
Korea Institute of Science and Technology Information and National Research Foundation of Korea (NRF), Republic of Korea;
Ministry of Education and Scientific Research, Institute of Atomic Physics, Ministry of Research and Innovation and Institute of Atomic Physics and University Politehnica of Bucharest, Romania;
Ministry of Education, Science, Research and Sport of the Slovak Republic, Slovakia;
National Research Foundation of South Africa, South Africa;
Swedish Research Council (VR) and Knut \& Alice Wallenberg Foundation (KAW), Sweden;
European Organization for Nuclear Research, Switzerland;
Suranaree University of Technology (SUT), National Science and Technology Development Agency (NSTDA), Thailand Science Research and Innovation (TSRI) and National Science, Research and Innovation Fund (NSRF), Thailand;
Turkish Energy, Nuclear and Mineral Research Agency (TENMAK), Turkey;
National Academy of  Sciences of Ukraine, Ukraine;
Science and Technology Facilities Council (STFC), United Kingdom;
National Science Foundation of the United States of America (NSF) and United States Department of Energy, Office of Nuclear Physics (DOE NP), United States of America.
In addition, individual groups or members have received support from:
Marie Sk\l{}odowska Curie, European Research Council, Strong 2020 - Horizon 2020 (grant nos. 950692, 824093, 896850), European Union;
Academy of Finland (Center of Excellence in Quark Matter) (grant nos. 346327, 346328), Finland;
Programa de Apoyos para la Superaci\'{o}n del Personal Acad\'{e}mico, UNAM, Mexico.

%TC:endignore
\end{acknowledgement}

\bibliographystyle{utphys} 
\bibliography{main}

\providecommand{\href}[2]{#2}\begingroup\raggedright\begin{thebibliography}{10}

\bibitem{navratil2016unified}
P.~Navr{\'a}til, S.~Quaglioni, G.~Hupin, C.~Romero-Redondo, and A.~Calci,
  ``Unified ab initio approaches to nuclear structure and reactions'',
  \href{http://dx.doi.org/10.1088/0031-8949/91/5/053002}{{\em Phys. Scripta}
  {\bfseries 91} (2016) 053002},
  \href{http://arxiv.org/abs/1601.03765}{{\ttfamily arXiv:1601.03765
  [nucl-th]}}.

\bibitem{doi:10.1146/annurev-nucl-102313-025446}
K.~Hebeler, J.~Holt, J.~Menéndez, and A.~Schwenk, ``Nuclear forces and their
  impact on neutron-rich nuclei and neutron-rich matter'',
  \href{http://dx.doi.org/10.1146/annurev-nucl-102313-025446}{{\em Ann. Rev.
  Nucl. Part. Sci.} {\bfseries 65} (2015) 457--484},
  \href{http://arxiv.org/abs/1508.06893}{{\ttfamily arXiv:1508.06893
  [nucl-th]}}.

\bibitem{Duer:2022ehf}
M.~Duer {\em et~al.}, ``{Observation of a correlated free four-neutron
  system}'', \href{http://dx.doi.org/10.1038/s41586-022-04827-6}{{\em Nature}
  {\bfseries 606} (2022) 678--682}.

\bibitem{RevModPhys.81.1773}
E.~Epelbaum, H.-W. Hammer, and U.-G. Mei\ss{}ner, ``Modern theory of nuclear
  forces'', \href{http://dx.doi.org/10.1103/RevModPhys.81.1773}{{\em Rev. Mod.
  Phys.} {\bfseries 81} (2009) 1773--1825},
  \href{http://arxiv.org/abs/0811.1338}{{\ttfamily arXiv:0811.1338 [nucl-th]}}.

\bibitem{barrett2013ab}
B.~R. Barrett, P.~Navr{\'a}til, and J.~P. Vary, ``Ab initio no core shell
  model'', \href{http://dx.doi.org/10.1016/j.ppnp.2012.10.003}{{\em Prog. Part.
  Nucl. Phys.} {\bfseries 69} (2013) 131--181}.

\bibitem{hagen2012evolution}
G.~Hagen, M.~Hjorth-Jensen, G.~Jansen, R.~Machleidt, and T.~Papenbrock,
  ``Evolution of shell structure in neutron-rich calcium isotopes'',
  \href{http://dx.doi.org/10.1103/PhysRevLett.109.032502}{{\em Phys. Rev.
  Lett.} {\bfseries 109} (2012) 032502},
  \href{http://arxiv.org/abs/1204.3612}{{\ttfamily arXiv:1204.3612 [nucl-th]}}.

\bibitem{hagen2014coupled}
G.~Hagen, T.~Papenbrock, M.~Hjorth-Jensen, and D.~J. Dean, ``Coupled-cluster
  computations of atomic nuclei'',
  \href{http://dx.doi.org/10.1088/0034-4885/77/9/096302}{{\em Rept. Prog.
  Phys.} {\bfseries 77} (2014) 096302},
  \href{http://arxiv.org/abs/1312.7872}{{\ttfamily arXiv:1312.7872 [nucl-th]}}.

\bibitem{soma2013ab}
V.~Soma, C.~Barbieri, and T.~Duguet, ``{Ab initio Gorkov-Green's function
  calculations of open-shell nuclei}'',
  \href{http://dx.doi.org/10.1103/PhysRevC.87.011303}{{\em Phys. Rev. C}
  {\bfseries 87} (2013) 011303},
  \href{http://arxiv.org/abs/1208.2472}{{\ttfamily arXiv:1208.2472 [nucl-th]}}.

\bibitem{roth2011similarity}
R.~Roth, J.~Langhammer, A.~Calci, S.~Binder, and P.~Navr{\'a}til,
  ``{Similarity-Transformed Chiral N N + 3 N Interactions for the Ab Initio
  Description of $^{12}$C and $^{16}$O}'',
  \href{http://dx.doi.org/10.1103/PhysRevLett.107.072501}{{\em Phys. Rev.
  Lett.} {\bfseries 107} (2011) 072501},
  \href{http://arxiv.org/abs/1105.3173}{{\ttfamily arXiv:1105.3173 [nucl-th]}}.

\bibitem{stroberg2019nonempirical}
S.~R. Stroberg, H.~Hergert, S.~K. Bogner, and J.~D. Holt, ``Nonempirical
  interactions for the nuclear shell model: an update'',
  \href{http://dx.doi.org/10.1146/annurev-nucl-101917-021120}{{\em Ann. Rev.
  Nucl. Part. Sci.} {\bfseries 69} (2019) 307--362},
  \href{http://arxiv.org/abs/1902.06154}{{\ttfamily arXiv:1902.06154
  [nucl-th]}}.

\bibitem{carlson2015quantum}
J.~Carlson, S.~Gandolfi, F.~Pederiva, S.~C. Pieper, R.~Schiavilla, K.~Schmidt,
  and R.~B. Wiringa, ``{Quantum Monte Carlo methods for nuclear physics}'',
  \href{http://dx.doi.org/10.1103/RevModPhys.87.1067}{{\em Rev. Mod. Phys.}
  {\bfseries 87} (2015) 1067}, \href{http://arxiv.org/abs/1412.3081}{{\ttfamily
  arXiv:1412.3081 [nucl-th]}}.

\bibitem{RevModPhys.85.197}
H.-W. Hammer, A.~Nogga, and A.~Schwenk, ``{Three-body forces: From cold atoms
  to nuclei}'', \href{http://dx.doi.org/10.1103/RevModPhys.85.197}{{\em Rev.
  Mod. Phys.} {\bfseries 85} (2013) 197},
  \href{http://arxiv.org/abs/1210.4273}{{\ttfamily arXiv:1210.4273 [nucl-th]}}.

\bibitem{Kievsky_2008}
A.~Kievsky, S.~Rosati, M.~Viviani, L.~E. Marcucci, and L.~Girlanda, ``A
  high-precision variational approach to three- and four-nucleon bound and
  zero-energy scattering states'',
  \href{http://dx.doi.org/10.1088/0954-3899/35/6/063101}{{\em J. Phys. G}
  {\bfseries 35} (2008) 063101},
  \href{http://arxiv.org/abs/0805.4688}{{\ttfamily arXiv:0805.4688 [nucl-th]}}.

\bibitem{STAR:2017gxa}
{\bfseries STAR} Collaboration, L.~Adamczyk {\em et~al.}, ``{Measurement of the
  $^3_{\Lambda}$H lifetime in Au+Au collisions at the BNL Relativistic Heavy
  Ion Collider}'', \href{http://dx.doi.org/10.1103/PhysRevC.97.054909}{{\em
  Phys. Rev. C} {\bfseries 97} (2018) 054909},
  \href{http://arxiv.org/abs/1710.00436}{{\ttfamily arXiv:1710.00436
  [nucl-ex]}}.

\bibitem{ALICE:2019vlx}
{\bfseries ALICE} Collaboration, S.~Acharya {\em et~al.},
  ``{$^3_\Lambda\mathrm{H}$ and $^3_{\bar{\Lambda}}\mathrm{\overline{H}}$
  lifetime measurement in Pb-Pb collisions at $\sqrt{s_{\mathrm{NN}}} = $ 5.02
  TeV via two-body decay}'',
  \href{http://dx.doi.org/10.1016/j.physletb.2019.134905}{{\em Phys. Lett. B}
  {\bfseries 797} (2019) 134905},
  \href{http://arxiv.org/abs/1907.06906}{{\ttfamily arXiv:1907.06906
  [nucl-ex]}}.

\bibitem{STAR:2019wjm}
{\bfseries STAR} Collaboration, J.~Adam {\em et~al.}, ``{Measurement of the
  mass difference and the binding energy of the hypertriton and
  antihypertriton}'', \href{http://dx.doi.org/10.1038/s41567-020-0799-7}{{\em
  Nature Phys.} {\bfseries 16} (2020) 409--412},
  \href{http://arxiv.org/abs/1904.10520}{{\ttfamily arXiv:1904.10520
  [hep-ex]}}.

\bibitem{ALargeIonColliderExperiment:2021puh}
{\bfseries ALICE} Collaboration, S.~Acharya {\em et~al.}, ``{Hypertriton
  Production in p-Pb Collisions at $\sqrt {s_\mathrm{NN}}$=5.02\,\,TeV}'',
  \href{http://dx.doi.org/10.1103/PhysRevLett.128.252003}{{\em Phys. Rev.
  Lett.} {\bfseries 128} (2022) 252003},
  \href{http://arxiv.org/abs/2107.10627}{{\ttfamily arXiv:2107.10627
  [nucl-ex]}}.

\bibitem{10.1143/PTP.103.929}
H.~Nemura, Y.~Suzuki, Y.~Fujiwara, and C.~Nakamoto, ``{Study of Light
  $\Lambda$- and $\Lambda\Lambda$-Hypernuclei with the Stochastic Variational
  Method and Effective $\Lambda$N Potentials}'',
  \href{http://dx.doi.org/10.1143/PTP.103.929}{{\em Prog. Theor. Phys.}
  {\bfseries 103} (2000) 929--958},
  \href{http://arxiv.org/abs/nucl-th/9912065}{{\ttfamily
  arXiv:nucl-th/9912065}}.

\bibitem{PhysRevC.100.034002}
F.~Hildenbrand and H.-W. Hammer, ``Three-body hypernuclei in pionless effective
  field theory'', \href{http://dx.doi.org/10.1103/PhysRevC.100.034002}{{\em
  Phys. Rev. C} {\bfseries 100} (2019) 034002},
  \href{http://arxiv.org/abs/1904.05818}{{\ttfamily arXiv:1904.05818
  [nucl-th]}}. [Erratum: Phys.Rev.C 102, 039901 (2020)].

\bibitem{Feliciello:2015dua}
A.~Feliciello and T.~Nagae, ``{Experimental review of hypernuclear physics:
  recent achievements and future perspectives}'',
  \href{http://dx.doi.org/10.1088/0034-4885/78/9/096301}{{\em Rept. Prog.
  Phys.} {\bfseries 78} (2015) 096301}.

\bibitem{PhysRevC.53.1210}
T.~Hasegawa {\em et~al.}, ``Spectroscopic study of
  $_{\mathrm{\ensuremath{\Lambda}}}^{10}\mathrm{B}$,
  $_{\mathrm{\ensuremath{\Lambda}}}^{12}\mathrm{C}$,
  $_{\mathrm{\ensuremath{\Lambda}}}^{28}\mathrm{Si}$,
  $_{\mathrm{\ensuremath{\Lambda}}}^{89}\mathrm{Y}$,
  $_{\mathrm{\ensuremath{\Lambda}}}^{139}\mathrm{La}$, and
  $_{\mathrm{\ensuremath{\Lambda}}}^{208}\mathrm{Pb}$ by the
  (${\mathrm{\ensuremath{\pi}}}^{+}$,${\mathit{k}}^{+}$) reaction'',
  \href{http://dx.doi.org/10.1103/PhysRevC.53.1210}{{\em Phys. Rev. C}
  {\bfseries 53} (Mar, 1996) 1210--1220}.

\bibitem{agnello2011hypernuclear}
{\bfseries FINUDA} Collaboration, M.~Agnello {\em et~al.}, ``{Hypernuclear
  spectroscopy with K$^-$ at rest on $^7$Li, $^9$Be, $^{13}$C and $^{16}$O}'',
  \href{http://dx.doi.org/10.1016/j.physletb.2011.02.060}{{\em Phys. Lett. B}
  {\bfseries 698} (2011) 219--225},
  \href{http://arxiv.org/abs/1011.2695}{{\ttfamily arXiv:1011.2695 [nucl-ex]}}.

\bibitem{PhysRevLett.120.132505}
{\bfseries J-PARC E13} Collaboration, S.~B. Yang {\em et~al.}, ``{First
  Determination of the Level Structure of an $sd$-Shell Hypernucleus,
  $_{\mathrm{\ensuremath{\Lambda}}}^{19}\mathrm{F}$}'',
  \href{http://dx.doi.org/10.1103/PhysRevLett.120.132505}{{\em Phys. Rev.
  Lett.} {\bfseries 120} (2018) 132505},
  \href{http://arxiv.org/abs/1712.07608}{{\ttfamily arXiv:1712.07608
  [nucl-ex]}}.

\bibitem{PhysRevC.99.054309}
{\bfseries Jefferson Lab Hall A} Collaboration, F.~Garibaldi {\em et~al.},
  ``High-resolution hypernuclear spectroscopy at jefferson lab, hall a'',
  \href{http://dx.doi.org/10.1103/PhysRevC.99.054309}{{\em Phys. Rev. C}
  {\bfseries 99} (2019) 054309},
  \href{http://arxiv.org/abs/1807.09720}{{\ttfamily arXiv:1807.09720
  [nucl-ex]}}.

\bibitem{PhysRevC.88.041001}
{\bfseries HypHI} Collaboration, C.~Rappold {\em et~al.}, ``{Search for
  evidence of ${}_{\ensuremath{\Lambda}}^{\phantom{\rule{0.28em}{0ex}}3}n$ by
  observing $d+{\ensuremath{\pi}}^{\ensuremath{-}}$ and
  $t+{\ensuremath{\pi}}^{\ensuremath{-}}$ final states in the reaction of
  ${}^{6}$Li+${}^{12}$C at $2A$ GeV}'',
  \href{http://dx.doi.org/10.1103/PhysRevC.88.041001}{{\em Phys. Rev. C}
  {\bfseries 88} (Oct, 2013) 041001}.

\bibitem{Saito:2021gao}
T.~R. Saito {\em et~al.}, ``{New directions in hypernuclear physics}'',
  \href{http://dx.doi.org/10.1038/s42254-021-00371-w}{{\em Nature Rev. Phys.}
  {\bfseries 3} (2021) 803--813}.

\bibitem{Oertel:2016bki}
M.~Oertel, M.~Hempel, T.~Kl\"ahn, and S.~Typel, ``{Equations of state for
  supernovae and compact stars}'',
  \href{http://dx.doi.org/10.1103/RevModPhys.89.015007}{{\em Rev. Mod. Phys.}
  {\bfseries 89} (2017) 015007},
  \href{http://arxiv.org/abs/1610.03361}{{\ttfamily arXiv:1610.03361
  [astro-ph.HE]}}.

\bibitem{Tolos:2020aln}
L.~Tolos and L.~Fabbietti, ``{Strangeness in Nuclei and Neutron Stars}'',
  \href{http://dx.doi.org/10.1016/j.ppnp.2020.103770}{{\em Prog. Part. Nucl.
  Phys.} {\bfseries 112} (2020) 103770},
  \href{http://arxiv.org/abs/2002.09223}{{\ttfamily arXiv:2002.09223
  [nucl-ex]}}.

\bibitem{natureNS}
P.~Demorest, T.~Pennucci, S.~Ransom, M.~Roberts, and J.~Hessels, ``{Shapiro
  Delay Measurement of A Two Solar Mass Neutron Star}'',
  \href{http://dx.doi.org/10.1038/nature09466}{{\em Nature} {\bfseries 467}
  (2010) 1081--1083}, \href{http://arxiv.org/abs/1010.5788}{{\ttfamily
  arXiv:1010.5788 [astro-ph.HE]}}.

\bibitem{Antoniadis1233232}
J.~Antoniadis {\em et~al.}, ``A massive pulsar in a compact relativistic
  binary'', \href{http://dx.doi.org/10.1126/science.1233232}{{\em Science}
  {\bfseries 340} (2013) 6131},
  \href{http://arxiv.org/abs/1304.6875}{{\ttfamily arXiv:1304.6875
  [astro-ph.HE]}}.

\bibitem{Lonardoni:2014bwa}
D.~Lonardoni, A.~Lovato, S.~Gandolfi, and F.~Pederiva, ``{Hyperon Puzzle: Hints
  from Quantum Monte Carlo Calculations}'',
  \href{http://dx.doi.org/10.1103/PhysRevLett.114.092301}{{\em Phys. Rev.
  Lett.} {\bfseries 114} (2015) 092301},
  \href{http://arxiv.org/abs/1407.4448}{{\ttfamily arXiv:1407.4448 [nucl-th]}}.

\bibitem{wiedemann1999particle}
U.~A. Wiedemann and U.~Heinz, ``Particle interferometry for relativistic
  heavy-ion collisions'',
  \href{http://dx.doi.org/10.1016/S0370-1573(99)00032-0}{{\em Phys. Rept.}
  {\bfseries 319} (1999) 145--230},
  \href{http://arxiv.org/abs/nucl-th/9901094}{{\ttfamily
  arXiv:nucl-th/9901094}}.

\bibitem{heinz1999two}
U.~Heinz and B.~V. Jacak, ``Two-particle correlations in relativistic heavy-ion
  collisions'', \href{http://dx.doi.org/10.1146/annurev.nucl.49.1.529}{{\em
  Ann. Rev. Nucl. Part. Sci.} {\bfseries 49} (1999) 529--579},
  \href{http://arxiv.org/abs/nucl-th/9902020}{{\ttfamily
  arXiv:nucl-th/9902020}}.

\bibitem{lednicky2004correlation}
R.~Lednick\'y, ``Correlation femtoscopy of multiparticle processes'',
  \href{http://dx.doi.org/10.1134/1.1644010}{{\em Phys. Atom. Nucl.} {\bfseries
  67} (2004) 72--82}, \href{http://arxiv.org/abs/nucl-th/0305027}{{\ttfamily
  arXiv:nucl-th/0305027}}.

\bibitem{Lisa:2005dd}
M.~A. Lisa, S.~Pratt, R.~Soltz, and U.~Wiedemann, ``{Femtoscopy in relativistic
  heavy ion collisions}'',
  \href{http://dx.doi.org/10.1146/annurev.nucl.55.090704.151533}{{\em Ann. Rev.
  Nucl. Part. Sci.} {\bfseries 55} (2005) 357--402},
  \href{http://arxiv.org/abs/nucl-ex/0505014}{{\ttfamily
  arXiv:nucl-ex/0505014}}.

\bibitem{adamczyk2015lambda}
{\bfseries STAR} Collaboration, L.~Adamczyk {\em et~al.}, ``{$\Lambda$
  $\Lambda$ Correlation Function in Au+ Au Collisions at
  $\sqrt{s_{\mathrm{NN}}}$ = 200 GeV}'',
  \href{http://dx.doi.org/10.1103/PhysRevLett.114.022301}{{\em Phys. Rev.
  Lett.} {\bfseries 114} (2015) 022301},
  \href{http://arxiv.org/abs/1408.4360}{{\ttfamily arXiv:1408.4360 [nucl-ex]}}.

\bibitem{STAR:2015kha}
{\bfseries STAR} Collaboration, L.~Adamczyk {\em et~al.}, ``{Measurement of
  Interaction between Antiprotons}'',
  \href{http://dx.doi.org/10.1038/nature15724}{{\em Nature} {\bfseries 527}
  (2015) 345--348}, \href{http://arxiv.org/abs/1507.07158}{{\ttfamily
  arXiv:1507.07158 [nucl-ex]}}.

\bibitem{STAR:2018uho}
{\bfseries STAR} Collaboration, J.~Adam {\em et~al.}, ``{The Proton-$\Omega$
  correlation function in Au+Au collisions at $\sqrt{s_\mathrm{NN}}$=200
  GeV}'', \href{http://dx.doi.org/10.1016/j.physletb.2019.01.055}{{\em Phys.
  Lett. B} {\bfseries 790} (2019) 490--497},
  \href{http://arxiv.org/abs/1808.02511}{{\ttfamily arXiv:1808.02511
  [hep-ex]}}.

\bibitem{ALICE:Run1}
{\bfseries ALICE} Collaboration, S.~Acharya {\em et~al.}, ``{p-p, p-$\Lambda$
  and $\Lambda$-$\Lambda$ correlations studied via femtoscopy in pp reactions
  at $\sqrt{s}=$7~\TeV}'',
  \href{http://dx.doi.org/10.1103/PhysRevC.99.024001}{{\em Phys. Rev. C}
  {\bfseries 99} (2019) 024001},
  \href{http://arxiv.org/abs/1805.12455}{{\ttfamily arXiv:1805.12455
  [nucl-ex]}}.

\bibitem{Acharya:2019bsa}
{\bfseries ALICE} Collaboration, S.~Acharya {\em et~al.}, ``{Scattering studies
  with low-energy kaon-proton femtoscopy in proton-proton collisions at the
  LHC}'', \href{http://dx.doi.org/10.1103/PhysRevLett.124.092301}{{\em Phys.
  Rev. Lett.} {\bfseries 124} (2020) 092301},
  \href{http://arxiv.org/abs/1905.13470}{{\ttfamily arXiv:1905.13470
  [nucl-ex]}}.

\bibitem{ALICE:pSig0}
{\bfseries ALICE} Collaboration, S.~Acharya {\em et~al.}, ``{Investigation of
  the p--$\Sigma^{0}$ interaction via femtoscopy in pp collisions}'',
  \href{http://dx.doi.org/10.1016/j.physletb.2020.135419}{{\em Phys. Lett. B}
  {\bfseries 805} (2020) 135419},
  \href{http://arxiv.org/abs/1910.14407}{{\ttfamily arXiv:1910.14407
  [nucl-ex]}}.

\bibitem{ALICE:LL}
{\bfseries ALICE} Collaboration, S.~Acharya {\em et~al.}, ``{Study of the
  $\Lambda$-$\Lambda$ interaction with femtoscopy correlations in pp and p-Pb
  collisions at the LHC}'',
  \href{http://dx.doi.org/10.1016/j.physletb.2019.134822}{{\em Phys. Lett. B}
  {\bfseries 797} (2019) 134822},
  \href{http://arxiv.org/abs/1905.07209}{{\ttfamily arXiv:1905.07209
  [nucl-ex]}}.

\bibitem{ALICE:pXi}
{\bfseries ALICE} Collaboration, S.~Acharya {\em et~al.}, ``{First Observation
  of an Attractive Interaction between a Proton and a Cascade Baryon}'',
  \href{http://dx.doi.org/10.1103/PhysRevLett.123.112002}{{\em Phys. Rev.
  Lett.} {\bfseries 123} (2019) 112002},
  \href{http://arxiv.org/abs/1904.12198}{{\ttfamily arXiv:1904.12198
  [nucl-ex]}}.

\bibitem{ALICE:pOmega}
{\bfseries ALICE} Collaboration, S.~Acharya {\em et~al.}, ``{Unveiling the
  strong interaction among hadrons at the LHC}'',
  \href{http://dx.doi.org/10.1038/s41586-020-3001-6}{{\em Nature} {\bfseries
  588} (2020) 232--238}, \href{http://arxiv.org/abs/2005.11495}{{\ttfamily
  arXiv:2005.11495}}.

\bibitem{ALICE:2021cpv}
{\bfseries ALICE} Collaboration, S.~Acharya {\em et~al.}, ``{Experimental
  Evidence for an Attractive p-$\phi$ Interaction}'',
  \href{http://dx.doi.org/10.1103/PhysRevLett.127.172301}{{\em Phys. Rev.
  Lett.} {\bfseries 127} (2021) 172301},
  \href{http://arxiv.org/abs/2105.05578}{{\ttfamily arXiv:2105.05578
  [nucl-ex]}}.

\bibitem{ALICE:2021cyj}
{\bfseries ALICE} Collaboration, S.~Acharya {\em et~al.}, ``{Investigating the
  role of strangeness in baryon\textendash{}antibaryon annihilation at the
  LHC}'', \href{http://dx.doi.org/10.1016/j.physletb.2022.137060}{{\em Phys.
  Lett. B} {\bfseries 829} (2022) 137060},
  \href{http://arxiv.org/abs/2105.05190}{{\ttfamily arXiv:2105.05190
  [nucl-ex]}}.

\bibitem{Fabbietti:2020bfg}
L.~Fabbietti, V.~M. Sarti, and O.~V. Doce, ``{Study of the strong interaction
  among hadrons with correlations at the LHC}'',
  \href{http://dx.doi.org/10.1146/annurev-nucl-102419-034438}{{\em Ann. Rev.
  Nucl. Part. Sci.} {\bfseries 71} (2021) 377--402},
  \href{http://arxiv.org/abs/2012.09806}{{\ttfamily arXiv:2012.09806
  [nucl-ex]}}.

\bibitem{Dhevan}
{\bfseries ALICE} Collaboration, B.~B. Abelev {\em et~al.}, ``{Two- and
  three-pion quantum statistics correlations in Pb-Pb collisions at
  $\sqrt{{s}_\mathrm{NN}}$ = 2.76 TeV at the CERN Large Hadron Collider}'',
  \href{http://dx.doi.org/10.1103/PhysRevC.89.024911}{{\em Phys. Rev. C}
  {\bfseries 89} (2014) 024911},
  \href{http://arxiv.org/abs/1310.7808}{{\ttfamily arXiv:1310.7808 [nucl-ex]}}.

\bibitem{PhysRevC.93.054908}
{\bfseries ALICE} Collaboration, J.~Adam {\em et~al.}, ``{Multipion
  Bose-Einstein correlations in pp, p-Pb, and Pb-Pb collisions at energies
  available at the CERN Large Hadron Collider}'',
  \href{http://dx.doi.org/10.1103/PhysRevC.93.054908}{{\em Phys. Rev. C}
  {\bfseries 93} (May, 2016) 054908},
  \href{http://arxiv.org/abs/1512.08902}{{\ttfamily arXiv:1512.08902}}.

\bibitem{Kubo}
R.~Kubo, ``{Generalized Cumulant Expansion Method}'',
  \href{http://dx.doi.org/10.1143/JPSJ.17.1100}{{\em J. Phys. Soc. Jpn.}
  {\bfseries 17} (1962) 1100--1120}.

\bibitem{DelGrande:2021mju}
R.~Del~Grande, L.~\v{S}erk\v{s}nyt\.{e}, L.~Fabbietti, V.~M. Sarti, and
  D.~Mihaylov, ``{A method to remove lower order contributions in
  multi-particle femtoscopic correlation functions}'',
  \href{http://dx.doi.org/10.1140/epjc/s10052-022-10209-z}{{\em Eur. Phys. J.
  C} {\bfseries 82} (2022) 244},
  \href{http://arxiv.org/abs/2107.10227}{{\ttfamily arXiv:2107.10227
  [nucl-th]}}.

\bibitem{ALICE}
{\bfseries ALICE} Collaboration, K.~Aamodt {\em et~al.}, ``{The ALICE
  experiment at the CERN LHC}'',
  \href{http://dx.doi.org/10.1088/1748-0221/3/08/S08002}{{\em J. Instr.}
  {\bfseries 3} (2008) S08002}.

\bibitem{ALICEperf}
{\bfseries ALICE} Collaboration, B.~Abelev {\em et~al.}, ``{Performance of the
  ALICE experiment at the CERN LHC}'',
  \href{http://dx.doi.org/10.1142/S0217751X14300440}{{\em Int. J. Mod. Phys.}
  {\bfseries A29} (2014) 1430044}.

\bibitem{Abbas:2013taa}
{\bfseries ALICE} Collaboration, E.~Abbas {\em et~al.}, ``{Performance of the
  ALICE VZERO system}'',
  \href{http://dx.doi.org/10.1088/1748-0221/8/10/P10016}{{\em JINST} {\bfseries
  8} (2013) P10016}, \href{http://arxiv.org/abs/1306.3130}{{\ttfamily
  arXiv:1306.3130 [nucl-ex]}}.

\bibitem{TPC}
J.~{Alme} {\em et~al.}, ``{The ALICE TPC, a large 3-dimensional tracking device
  with fast readout for ultra-high multiplicity events}'',
  \href{http://dx.doi.org/10.1016/j.nima.2010.04.042}{{\em Nucl. Instrum.
  Methods} {\bfseries A622} (2010) 316--367}.

\bibitem{TOF}
A.~Akindinov {\em et~al.}, ``{Performance of the ALICE Time-of-Flight detector
  at the LHC}'', \href{http://dx.doi.org/10.1140/epjp/i2013-13044-x}{{\em Eur.
  Phys. J. Plus} {\bfseries 128} (2013) 44}.

\bibitem{Acharya:2020dfb}
{\bfseries ALICE} Collaboration, S.~Acharya {\em et~al.}, ``{Search for a
  common baryon source in high-multiplicity pp collisions at the LHC}'',
  \href{http://dx.doi.org/10.1016/j.physletb.2020.135849}{{\em Phys. Lett. B}
  {\bfseries 811} (2020) 135849},
  \href{http://arxiv.org/abs/2004.08018}{{\ttfamily arXiv:2004.08018
  [nucl-ex]}}.

\bibitem{PDG}
{\bfseries Particle Data Group} Collaboration, M.~Tanabashi {\em et~al.},
  ``{Review of Particle Physics}'',
  \href{http://dx.doi.org/10.1103/PhysRevD.98.030001}{{\em Phys. Rev.}
  {\bfseries D98} (2018) 030001}.

\bibitem{Albrecht:1986me}
{\bfseries ARGUS} Collaboration, H.~Albrecht {\em et~al.}, ``{Observation of
  Octet and Decuplet Hyperons in $e^+ e^-$ Annihilation at 10-{GeV}
  Center-of-mass Energy}'',
  \href{http://dx.doi.org/10.1016/0370-2693(87)90988-9}{{\em Phys. Lett. B}
  {\bfseries 183} (1987) 419--424}.

\bibitem{Lednicky:1981su}
R.~Lednick\'y and V.~L. Lyuboshits, ``{Final State Interaction Effect on
  Pairing Correlations Between Particles with Small Relative Momenta}'', {\em
  Sov. J. Nucl. Phys.} {\bfseries 35} (1982) 770.

\bibitem{koonin1977proton}
S.~E. Koonin, ``Proton pictures of high-energy nuclear collisions'',
  \href{http://dx.doi.org/10.1016/0370-2693(77)90340-9}{{\em Phys. Lett. B}
  {\bfseries 70} (1977) 43--47}.

\bibitem{pratt1990detailed}
S.~Pratt, T.~Cs{\"o}rg{\H{o}}, and J.~Zim\'anyi, ``Detailed predictions for
  two-pion correlations in ultrarelativistic heavy-ion collisions'',
  \href{http://dx.doi.org/10.1103/PhysRevC.42.2646}{{\em Phys. Rev. C}
  {\bfseries 42} (1990) 2646}.

\bibitem{ALICE:2021njx}
{\bfseries ALICE} Collaboration, S.~Acharya {\em et~al.}, ``{Exploring the
  N$\Lambda$-N$\Sigma$ coupled system with high precision correlation
  techniques at the LHC}'',
  \href{http://dx.doi.org/10.1016/j.physletb.2022.137272}{{\em Phys. Lett. B}
  {\bfseries 833} (2022) 137272},
  \href{http://arxiv.org/abs/2104.04427}{{\ttfamily arXiv:2104.04427
  [nucl-ex]}}.

\bibitem{Niemann:2012gd}
P.~Niemann and H.~W. Hammer, ``{Pauli blocking effects and Cooper triples in
  three-component Fermi gases}'',
  \href{http://dx.doi.org/10.1103/PhysRevA.86.013628}{{\em Phys. Rev. A}
  {\bfseries 86} (2012) 013628},
  \href{http://arxiv.org/abs/1203.1824}{{\ttfamily arXiv:1203.1824
  [cond-mat.quant-gas]}}.

\bibitem{alt1993asymptotic}
E.~O. Alt and A.~M. Mukhamedzhanov, ``Asymptotic solution of the
  {S}chr{\"o}dinger equation for three charged particles'',
  \href{http://dx.doi.org/10.1103/PhysRevA.47.2004}{{\em Phys. Rev. A}
  {\bfseries 47} (1993) 2004--2022}.

\bibitem{Arndt:2007qn}
R.~A. Arndt, W.~J. Briscoe, I.~I. Strakovsky, and R.~L. Workman, ``{Updated
  analysis of NN elastic scattering to 3-GeV}'',
  \href{http://dx.doi.org/10.1103/PhysRevC.76.025209}{{\em Phys. Rev. C}
  {\bfseries 76} (2007) 025209},
  \href{http://arxiv.org/abs/0706.2195}{{\ttfamily arXiv:0706.2195 [nucl-th]}}.

\bibitem{NavarroPerez:2013usk}
R.~Navarro~P\'erez, J.~E. Amaro, and E.~Ruiz~Arriola, ``{Partial Wave Analysis
  of Nucleon-Nucleon Scattering below pion production threshold}'',
  \href{http://dx.doi.org/10.1103/PhysRevC.88.024002}{{\em Phys. Rev. C}
  {\bfseries 88} (2013) 024002},
  \href{http://arxiv.org/abs/1304.0895}{{\ttfamily arXiv:1304.0895 [nucl-th]}}.
  [Erratum: Phys.Rev.C 88, 069902 (2013)].

\bibitem{ALICE-pp-publicnote}
{\bfseries ALICE} Collaboration, ``{Future high-energy pp programme with
  ALICE}'', {\em ALICE-PUBLIC-2020-005, CERN-LHCC-2020-018, LHCC-G-179} (2020)
  . \url{https://cds.cern.ch/record/2724925}.

\bibitem{Sjostrand:2014zea}
T.~Sj\"ostrand, S.~Ask, J.~R. Christiansen, R.~Corke, N.~Desai, P.~Ilten,
  S.~Mrenna, S.~Prestel, C.~O. Rasmussen, and P.~Z. Skands, ``{An introduction
  to PYTHIA 8.2}'', \href{http://dx.doi.org/10.1016/j.cpc.2015.01.024}{{\em
  Comput. Phys. Commun.} {\bfseries 191} (2015) 159--177},
  \href{http://arxiv.org/abs/1410.3012}{{\ttfamily arXiv:1410.3012 [hep-ph]}}.

\bibitem{ALICEac2020}
{\bfseries ALICE} Collaboration, J.~Adam {\em et~al.}, ``{Insight into particle
  production mechanisms via angular correlations of identified particles in pp
  collisions at $\sqrt{s}$ = 7 TeV}'',
  \href{http://dx.doi.org/10.1140/epjc/s10052-017-5129-6}{{\em Eur. Phys. J. C}
  {\bfseries 77} (2017) 569}, \href{http://arxiv.org/abs/1612.08975}{{\ttfamily
  arXiv:1612.08975 [nucl-ex]}}. [Erratum: Eur.Phys.J.C 79, 998 (2019)].

\end{thebibliography}\endgroup

%%%%%%%%% appendix with author list
\newpage
\appendix

\section{Monte Carlo studies}
\label{app:MC}
\begin{figure}[!ht]
\centering
\includegraphics[width=0.32\textwidth]{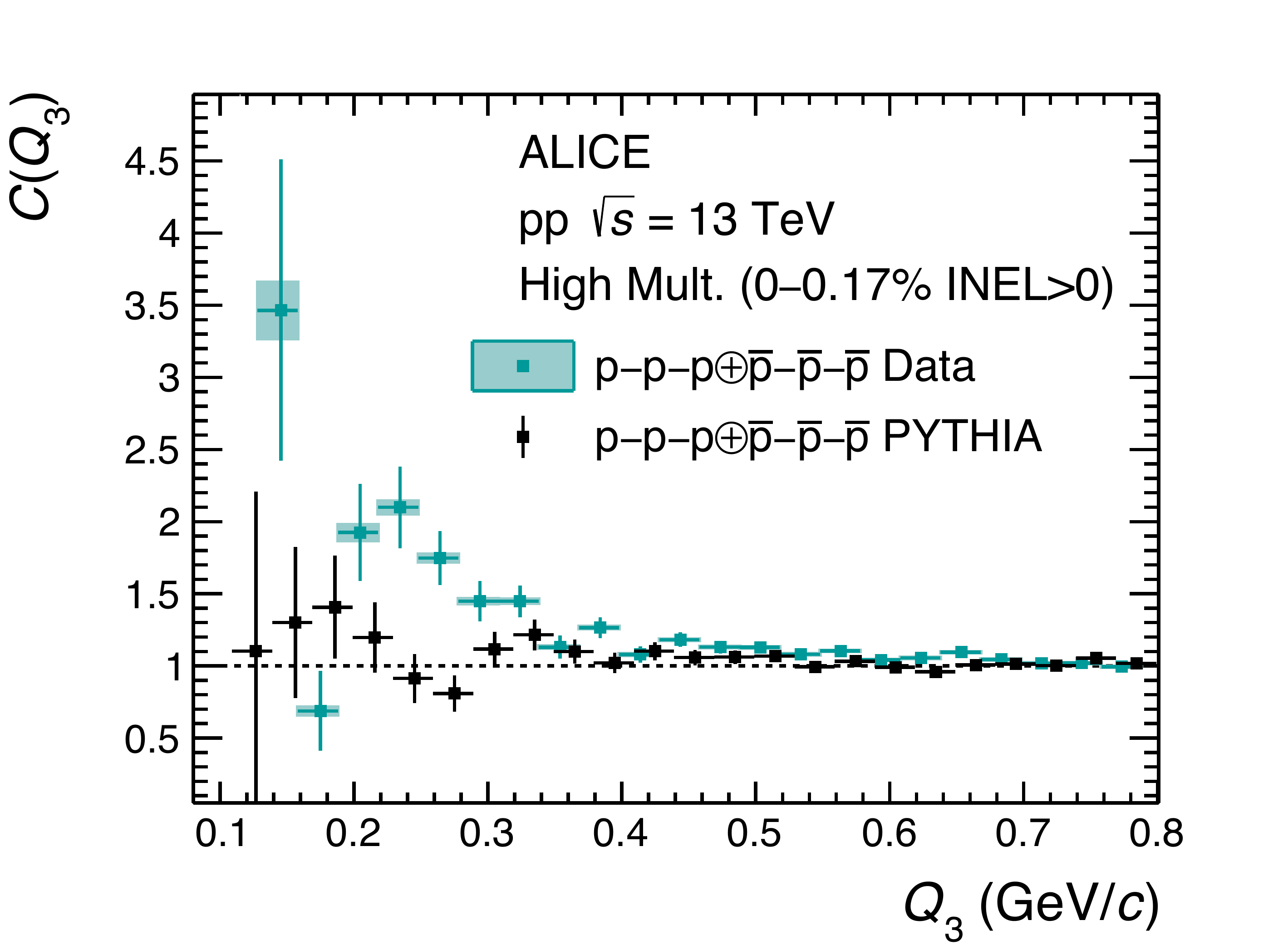}
\includegraphics[width=0.32\textwidth]{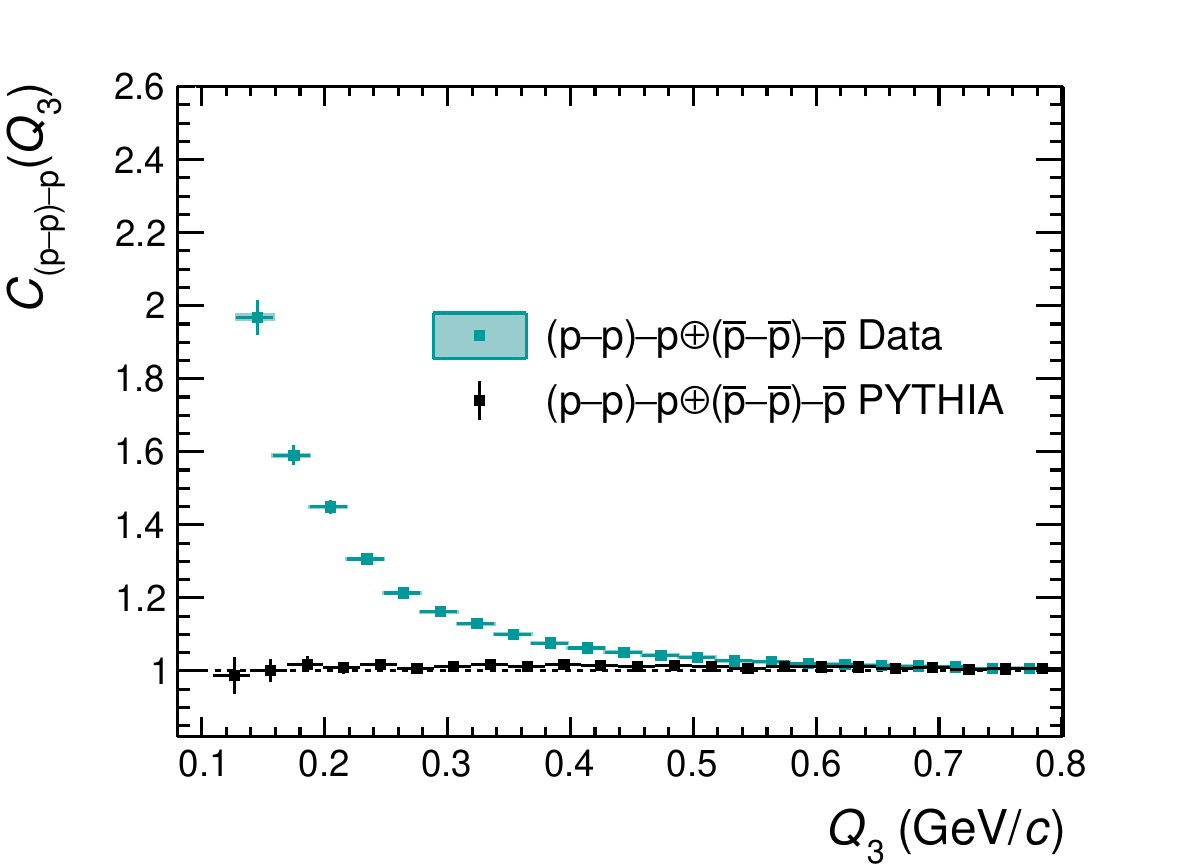}\\
\includegraphics[width=0.32\textwidth]{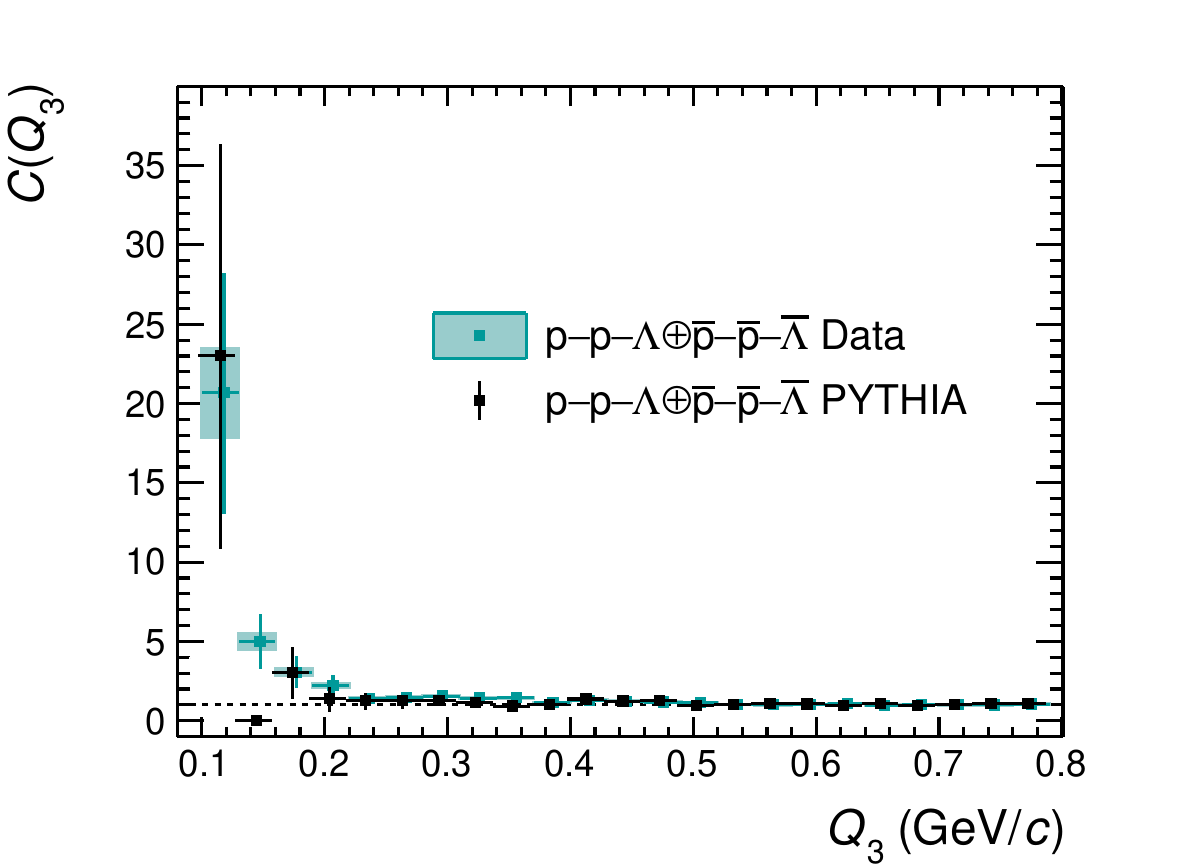}
\includegraphics[width=0.32\textwidth]{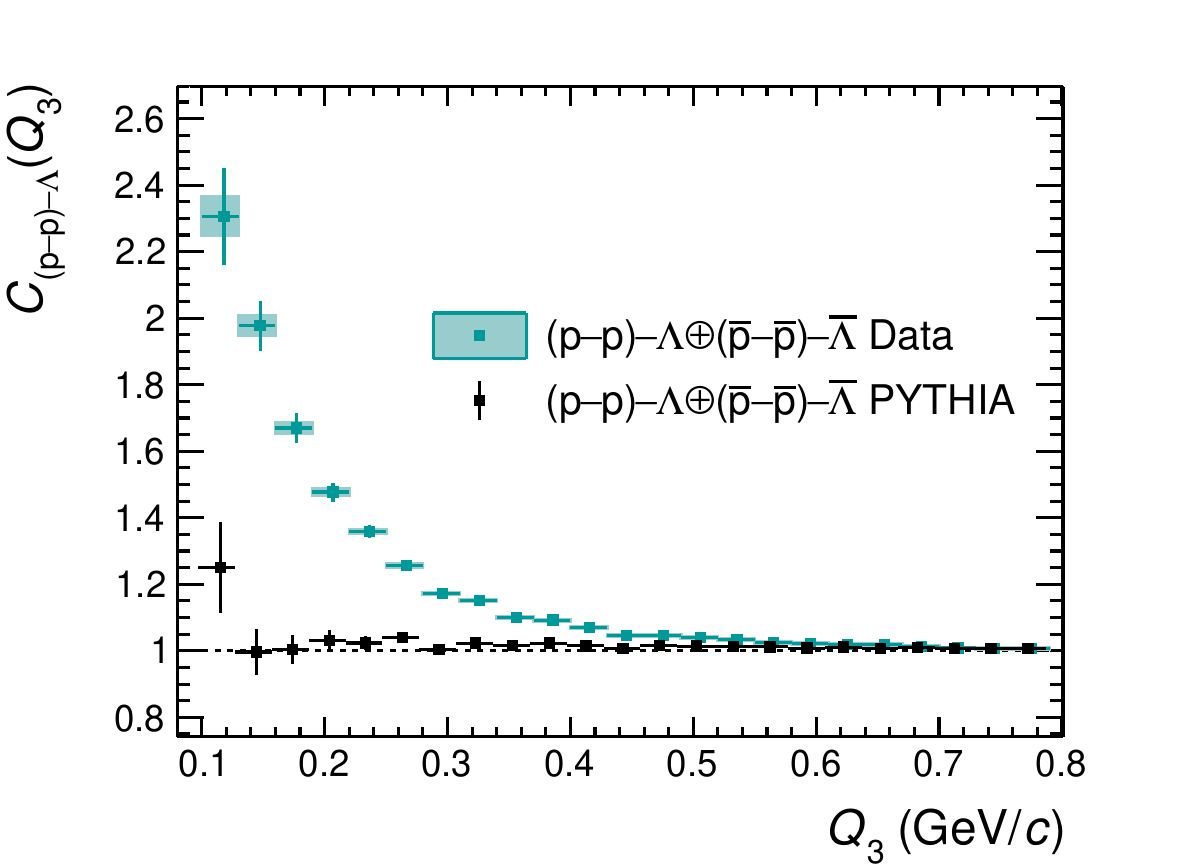}
\includegraphics[width=0.32\textwidth]{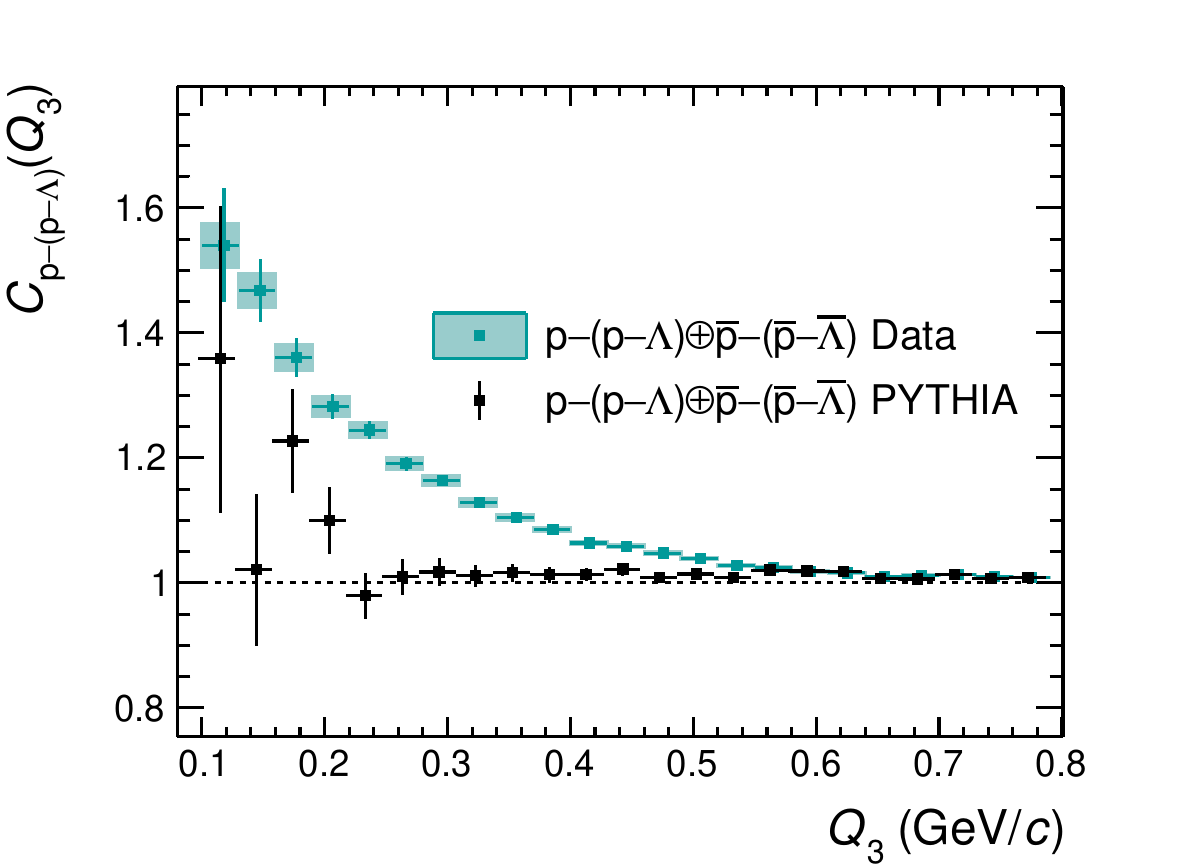}
\caption{The comparison between the correlation functions obtained from the measurements (green) and from the PYTHIA 8 event generator with a dedicated high-multiplicity selection to mimic the V0 high-multiplicity trigger (black).}
\label{fig:sigmainel_results}
\end{figure}

One of the benchmarks of the presented analysis consists in verifying that the three-particle correlations obtained from PYTHIA 8~\cite{Sjostrand:2014zea} (Monash 2013 Tune) simulations do not show any significant deviation from unity. Indeed, no FSIs -- either two- or three-particle -- are included in the simulated and reconstructed Monte Carlo data using the PYTHIA 8 event generator for pp collisions at $\sqrt{s} = \, 13$ TeV, thus Monte Carlo can neither be used to estimate the lower-order contributions (two-body correlations) nor genuine three-body effects. Figure~\ref{fig:sigmainel_results} shows the comparison between the measured and simulated correlation functions as a function of $Q_3$, where the simulation includes a dedicated high-multiplicity selection to mimic the V0 high-multiplicity trigger in the real data. The green symbols represent the experimental data while the black symbols refer to the simulation.  The simulated correlation functions are consistent with unity for the entire $Q_3< 0.8 $ GeV/$c$ range, showing that there are no effects caused by the track reconstruction in the detector, as well as no sign of mini-jets contribution (see more details in Ref.~\cite{ALICE:2021cyj}) and that the energy and momentum conservation effects are eventually present at larger values of $Q_3$ that are not relevant for the studies carried out in this work. Also the simulations of the  lower-order contributions display the same behaviour.

\section{Mixed-charge correlation studies}
\label{app:MixedChargeCumulant}

\begin{figure}[ht]
\centering
\includegraphics[width=0.48\textwidth]{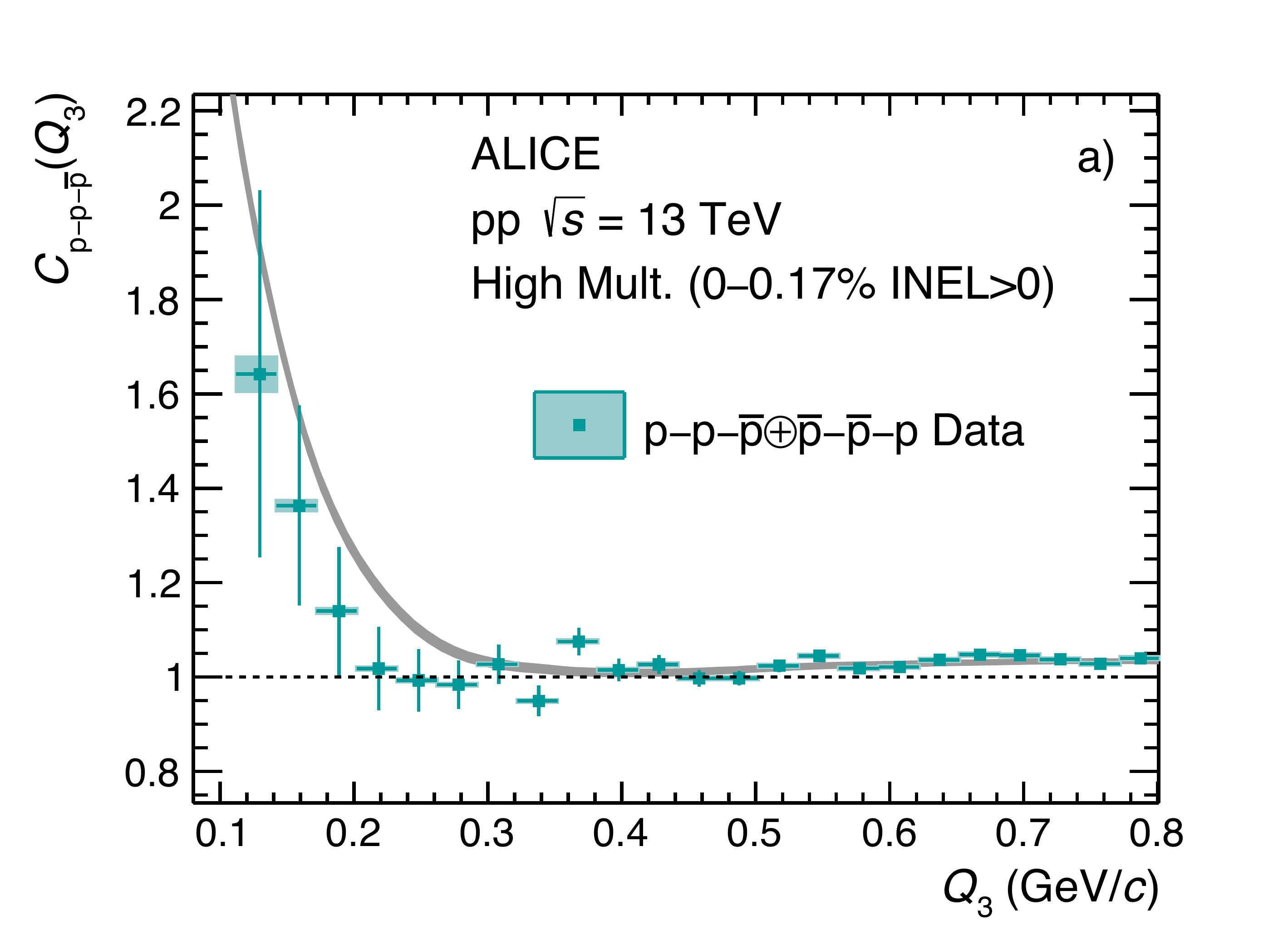}\\
\includegraphics[width=0.48\textwidth]{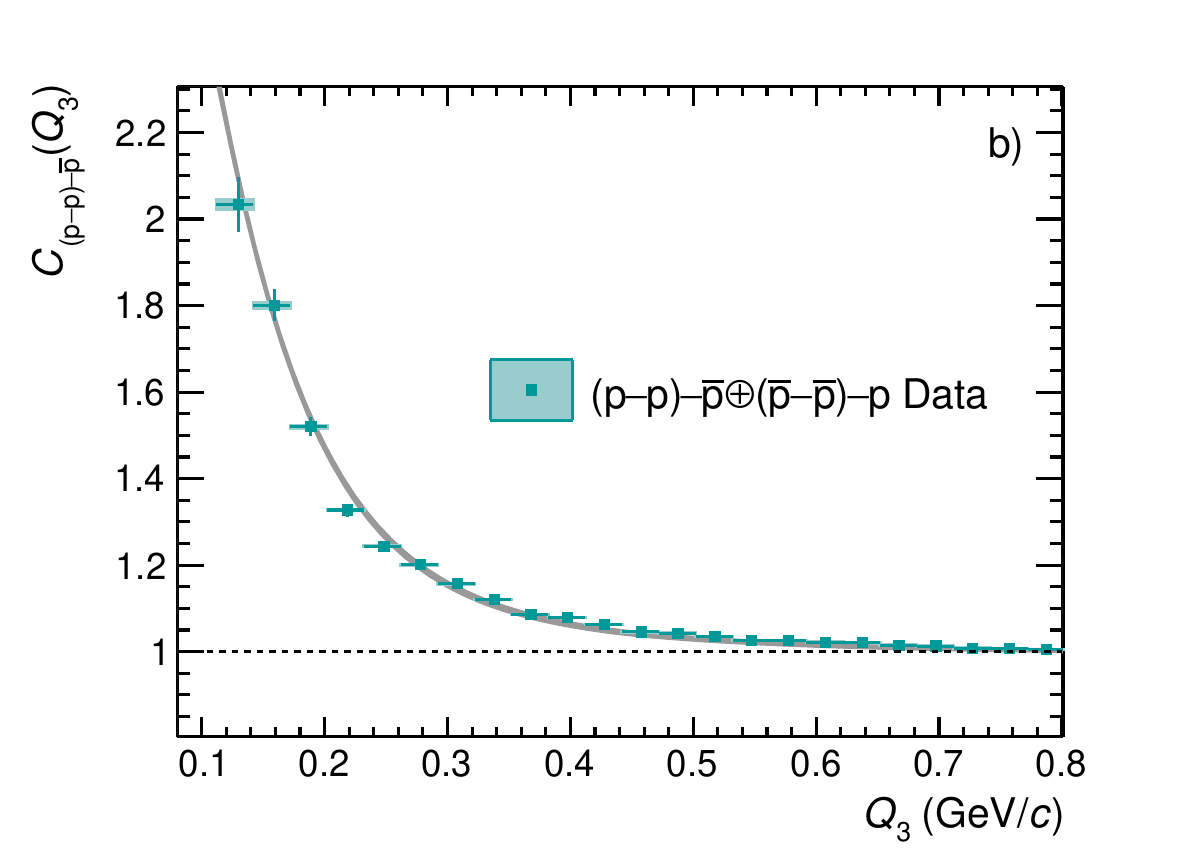}
\includegraphics[width=0.48\textwidth]{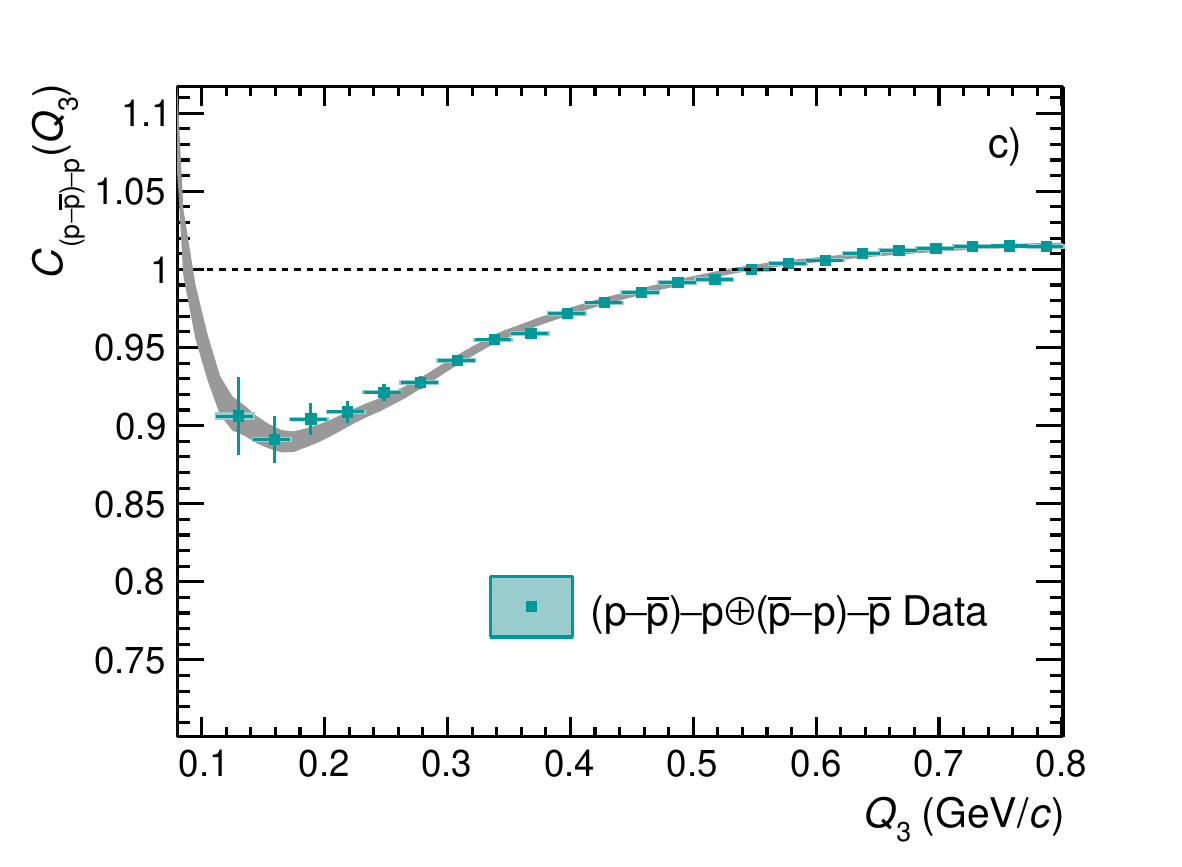}
\caption{Panel a) shows the correlation function for \pppbarCF triplets (green data points) and the total lower order contributions (grey band). Panels b) and c) show the \ppspbarCF and \pppbarsCF lower-order contributions to the measured \pppbarCF correlation function. The error bars and the boxes represent the statistical  and systematic uncertainties, respectively. The grey band includes systematic and statistical uncertainties summed in quadrature obtained from the projector method.}
\label{fig:mixedcharge_results}
\end{figure}

An additional benchmark for the \pppCF cumulant result and the measured deviation from zero has been considered by studying the \pppbarCF triplets. Identical event and track selection criteria and systematic variations as those employed for the \pppCF analysis have been used. The rejection in the $\Delta \eta$--$\Delta \varphi$ plane (see Section~\ref{3corr} for the details) has been applied only for same-charge pairs in the triplet. The obtained correlation functions for \pppbarCF, \ppspbarCF and \pppbarsCF triplets with the corresponding statistical and systematic uncertainties are shown in panels a), b) and c) of  Fig.~\ref{fig:mixedcharge_results}, respectively. The grey bands are obtained using the projector method following the procedure described in Section~\ref{sec:projector}.
The \ppspbarCF correlation functions obtained with the data-driven approach and the projector method (panel b) in Fig.~\ref{fig:mixedcharge_results} are in agreement with the \ppspCF results (panel a) in Fig.~\ref{fig:2bodyseparate}. This is consistent with the expectation from Eqs.~\ref{eq:C3ij} and~\ref{projector}, since such correlation functions depend only on the correlation of the two particles in the same event and on the mass of the uncorrelated particle, which are identical in the \ppspbarCF and \ppspCF systems. The \pppbarsCF correlation function reflects the interplay of FSIs and non-femtoscopic correlations measured in the study of \ppbarCF pairs~\cite{ALICE:2021cyj}. The grey band in panel c) of Fig.~\ref{fig:mixedcharge_results} is obtained by using as input of Eq.~\ref{eq:C3ij} the correlation function $C(k^*)$ of \ppbarCF pairs emitted in 
\pppbarCF triplets with $Q_3 < 1$ GeV/$c$. 
Such selection is performed 
in order to use \ppbarCF pairs produced in similar shape events as those where low $Q_3$ triplets are found. The requirement is necessary for \ppbarCF correlations due to the mini-jet contribution~\cite{ALICE:2021cyj} which is instead not present in p--$\Lambda$ and p--p (see e.g.~\cite{ALICEac2020}).  
The systematics induced by this additional selection are considered in the uncertainties.

The \pppbarCF cumulant is obtained using the projector method and it is shown in Fig.~\ref{fig:cumulants} by red circular symbols. The result is compatible with zero within the statistical and systematic uncertainties, demonstrating that strong-interaction as well as Coulomb effects on three-particle level are not statistically significant in \pppbarCF.   

\section{Feed-down contributions}
\label{sec:AppendixB}

The measured femtoscopic correlations do not originate only from correctly identified primary particles but they include as well contributions from  misidentified particles and feed-down particles originating from weakly decaying hadrons. In case of two-body femtoscopy, the decomposition method explained in Section~\ref{sec:decomposition} and the Eq.~\ref{eq:Ck_lambda} can be used to account for such impurities and feed-down effects. \\
This method can be extended to the three-particle case. The total data sample $X$ contains the particles which stem from feed-down $X_F$, misidentified particles $X_M$ and the correctly identified primary particles $X_0$. Both feed-down and misidentified particles can originate from different channels and the contributions can be expressed as 

\begin{equation}
X_{F}=\sum_{i=1}^{N_{F}} X_{i},
\end{equation}
\begin{equation}
X_{M}=\sum_{i=N_F+1}^{N_{F} + N_{M}} X_{i},
\end{equation}
where $N_F$ and $N_M$ are the number of feed-down and misidentification contributions.
The purity is the fraction of correctly identified particles to the total number of particles in the data sample and can be defined as
\begin{equation}
\mathcal{P}(X)=\left(X_{0}+X_{F}\right) / X.
\end{equation}
The correctly identified particles can stem from the decays of particles and for this purpose the channel fraction $f(X_i)$ is defined as
\begin{equation}
f(X_i)=X_i /\left(X_{0}+X_{F}\right).
\end{equation}
The fraction of specific channel in the whole data sample then can be written as
\begin{equation}
P(X_i)=\mathcal{P}(X_i) f(X_i)=\frac{X_i}{X}.
\label{eq:theGuyWithoutAName}
\end{equation}
 The correlation function for three particles can be written as
\begin{equation}
C(XYZ) = \frac{N(XYZ)}{M(XYZ)},
\end{equation}
where $N$ and $M$ are the yields of $XYZ$ triplet in same and mixed events, respectively. Using the identities introduced before one can write

\begin{equation}
N(X Y Z)=N\left(\sum_{i, j,k} X_{i} Y_{j} Z_{k}\right)=\sum_{i, j,k} N\left(X_{i} Y_{j}Z_{k}\right) ,
\end{equation}
\begin{equation}
M(X Y Z)=M\left(\sum_{i, j,k} X_{i} Y_{j}Y_{k}\right)=\sum_{i, j,k} M\left(X_{i} Y_{j}Z_{k}\right).
\end{equation}

The correlation function then can be rewritten as
\begin{equation}
\begin{aligned}
C(X Y Z) &=\frac{\sum_{i, j,k} N\left(X_{i} Y_{j}Z_{k}\right)}{M(X Y Z)}=\sum_{i, j,k} \frac{N\left(X_{i} Y_{j}Z_{k}\right)}{M(X YZ)} \frac{M\left(X_{i} Y_{j} Z_{k}\right)}{M\left(X_{i} Y_{j}Z_{k}\right)}=\\
&=\sum_{i, j,k} \underbrace{\frac{N\left(X_{i} Y_{j}Z_{k}\right)}{M\left(X_{i} Y_{j}Z_{k}\right)}}_{C_{i, j,k}(X YZ)} \underbrace{\frac{M\left(X_{i} Y_{j}Z_{k}\right)}{M(X YZ)}}_{\lambda_{i, j,k}(X Y Z)}=\sum_{i, j,k} \lambda_{i, j,k}(X Y Z) C_{i, j,k}(X YZ),
\end{aligned}
\label{eq:derivelambdasCF3}
\end{equation}
where $C_{i, j,k}(X YZ)$ is the correlation function of the $i,j,k$-th channel of origin of the particles $X,Y,Z$ and the $ \lambda_{i, j,k}(X Y Z)$ is the weight for such correlation. This parameter depends only on the mixed event sample and can be related to previously introduced single particle quantities, channel fraction and purity, as follows
\begin{equation}
\begin{aligned}
\lambda_{i, j,k}(X YZ)&=\frac{M\left(X_{i} Y_{j} Z_{k}\right)}{M(X YZ)}\\
&=\frac{M\left(X_{i}\right)}{M(X)} \frac{M\left(Y_{j}\right)}{M(Y)} \frac{M\left(Z_{k}\right)}{M(Z)}=P\left(X_{i}\right) P\left(Y_{j}\right) P\left(Z_{k}\right)\\
 &=\mathcal{P}(X_i) f(X_i)\mathcal{P}(Y_j) f(Y_j)\mathcal{P}(Z_k) f(Z_k) \ .
\end{aligned}
\label{eq:lambda3}
\end{equation}

To study the lower-order contributions in the measured three-particle correlation function, one needs to define the decomposition for the measurement of two correlated particles. In such case, the origin of the third particle in the numerator does not matter as it is uncorrelated. Equation~\ref{eq:derivelambdasCF3} becomes
\begin{equation}
\begin{aligned}
C(X Y, Z) &=\frac{\sum_{i, j} N\left(X_{i} Y_{j},Z\right)}{M(X Y, Z)}=\sum_{i, j} \frac{N\left(X_{i} Y_{j},Z\right)}{M(X Y,Z)} \frac{M\left(X_{i} Y_{j}, Z\right)}{M\left(X_{i} Y_{j},Z\right)}=\\
&=\sum_{i, j} \underbrace{\frac{N\left(X_{i} Y_{j},Z\right)}{M\left(X_{i} Y_{j},Z\right)}}_{C_{i, j}(X Y,Z)} \underbrace{\frac{M\left(X_{i} Y_{j},Z\right)}{M(X Y,Z)}}_{\lambda_{i, j}(X Y, Z )}=\sum_{i, j} \lambda_{i, j}(X Y, Z) C_{i, j}(X Y,Z).
\end{aligned}
\label{eq:derivelambdasCF2}
\end{equation}
Here $C_{i, j}(X Y,Z)$ denotes the correlation function where the two correlated particles $X$ and $Y$ are from origin $i$ and $j$ respectively and $Z$ is from any origin.  
Here $\lambda_{i, j}(X Y, Z) $ is

\begin{equation}
\begin{aligned}
\lambda_{i, j}(X Y,Z)&=\frac{M\left(X_{i} Y_{j}, Z\right)}{M(X Y,Z)}\\
&=\frac{M\left(X_{i}\right)}{M(X)} \frac{M\left(Y_{j}\right)}{M(Y)} \frac{M\left(Z\right)}{M(Z)}=P\left(X_{i}\right) P\left(Y_{j}\right) \cdot 1\\
 &=\mathcal{P}(X_i) f(X_i)\mathcal{P}(Y_j) f(Y_j) \ .
\end{aligned}
\label{eq:lambda2}
\end{equation}

This notation is valid only if one wants to study the $(X-Y)-Z$ correlation. To obtain the cumulant of the primary particles which were identified correctly, one has to subtract the lower-order correlations, such as $(X-Y)-Z$, from the three-particle correlation. For this purpose, the Eq.~\ref{eq:derivelambdasCF2} must be rewritten. As previously explained, in case of $(X-Y)-Z$ correlation, the origin of the third uncorrelated particle is not important, which means that $C(X Y,Z_l)=C(X Y,Z_m)$, where $Z_l$ and $Z_m$ have different origin. As previously shown, the fraction of particles $Z_l$ in the whole data sample is $P(Z_l)$, which can be as well expressed with the $\lambda$ parameter of one particle $\lambda_l(Z)$. Using the property $1 = \sum_{k}\lambda_k(Z)$ of the $\lambda$ parameters in Eq.~\ref{eq:lambda2}, one can write

\begin{equation}
\begin{aligned}
\lambda_{i, j}(X Y, Z) = \lambda_{i, j}(X Y, Z) \sum_k \lambda_{k}(Z) = \sum_k \lambda_{i, j}(X Y, Z)  \lambda_{k}(Z) = \sum_k \lambda_{i, j,k}(X Y Z) \ ,
\end{aligned}
\label{eq:sexyLambda}
\end{equation}

and Eq.~\ref{eq:derivelambdasCF2} can be rewritten as 
\begin{equation}
\begin{aligned}
C(X Y, Z) &=\sum_{i, j} \lambda_{i, j}(X Y, Z) C_{i, j}(X Y,Z)\\
&=\sum_{i, j}  \lambda_{i, j}(X Y, Z) \sum_{k} \lambda_{k}(Z)C_{i,j}(XY, Z) \\
&= \sum_{i, j}  \sum_{k}  \lambda_{i, j}(X Y, Z) \lambda_{k}(Z)C_{i,j}(X Y, Z)\\
&= \sum_{i, j, k}  \lambda_{i, j, k}(X Y Z) C_{i,j}(X Y, Z).
\end{aligned}
\label{eq:derivelambdasCF23}
\end{equation}

In the following, the above formalism is used to obtain the correctly identified primary particle cumulant. Starting with the cumulant expression

\begin{equation}
\begin{aligned}
c(XYZ) &= C(XYZ)-C(XY,Z)-C(XZ,Y)-C(ZY,X)+2 \\
&= \sum_{i, j,k} \lambda_{i, j,k}(X Y Z) C_{i, j,k}(X YZ) -  \sum_{i, j, k}   \lambda_{i, j,k}(X Y Z) C_{i,j}(X Y, Z) \\
&-  \sum_{i, j, k}   \lambda_{i, j,k}(X Y Z) C_{i,k}(X Z, Y)-  \sum_{i, j, k}   \lambda_{i, j,k}(X Y Z) C_{k,j}(Z Y, X) +2,
\end{aligned}
\label{eq:derivecf32}
\end{equation}
the correctly identified primary particle correlations can be isolated from the rest as follows

\begin{equation}
\begin{aligned}
c(XYZ) &=\left( \lambda_{X_0 Y_0 Z_0}(X YZ) C_{X_0 Y_0 Z_0}(X YZ) + \sum_{i, j,k\neq (X_0 Y_0 Z_0)} \lambda_{i, j,k}(X Y Z) C_{i, j,k}(X YZ)\right) \\
&- \left( \lambda_{X_0 Y_0 Z_0}(X Y,Z) C_{X_0 Y_0 }(X Y,Z) + \sum_{i, j, k\neq (X_0 Y_0 Z_0)}   \lambda_{i, j,k}(X Y Z) C_{i,j}(X Y, Z)\right)\\
&- \left( \lambda_{X_0 Y_0 Z_0}(X Z,Y) C_{X_0 Z_0 }(X Z, Y) + \sum_{i, j, k\neq (X_0 Y_0 Z_0)}   \lambda_{i, j,k}(X Y Z) C_{i,k}(X Z, Y)\right) \\
&- \left( \lambda_{X_0 Y_0 Z_0}(ZY, X) C_{ Z_0 Y_0}(ZY, X) + \sum_{i, j, k\neq (X_0 Y_0 Z_0)}   \lambda_{i, j,k}(X Y Z) C_{k,j}(ZY, X)\right) +2 \ .
\end{aligned}
\label{eq:derivecf32r}
\end{equation}

Written in such a way, one can already see that the cumulant of the measured correlation function can be expressed as the sum of the correctly identified primary particle cumulant and the cumulant of the rest of the contributions as follows

\begin{equation}
\begin{aligned}
c(XYZ) &= \lambda_{X_0 Y_0 Z_0}(X YZ) C_{X_0 Y_0 Z_0}(X YZ) -  \lambda_{X_0 Y_0 Z_0}(X Y,Z) C_{X_0 Y_0 }(X Y,Z)\\
&- \lambda_{X_0 Y_0 Z_0}(X Z,Y) C_{X_0 Z_0 }(X Z, Y) - \lambda_{X_0 Y_0 Z_0}(ZY, X) C_{ Z_0 Y_0}(ZY, X) \\ &+\sum_{i, j,k\neq (X_0 Y_0 Z_0)} \lambda_{i, j,k}(X Y Z) C_{i, j,k}(X YZ) - \sum_{i, j, k\neq (X_0 Y_0 Z_0)}   \lambda_{i, j,k}(X Y Z) C_{i,j}(X Y, Z)\\
&- \sum_{i, j, k\neq (X_0 Y_0 Z_0)}   \lambda_{i, j,k}(X Y Z) C_{i,k}(X Z, Y)- \sum_{i, j, k\neq (X_0 Y_0 Z_0)}   \lambda_{i, j,k}(X Y Z) C_{k,j}(ZY, X) +2 \\
& = \lambda_{X_0 Y_0 Z_0}(X YZ) \left(C_{X_0 Y_0 Z_0}(X YZ) -   C_{X_0 Y_0 }(X Y,Z)- C_{X_0 Z_0 }(X Z, Y) -  C_{ Z_0 Y_0}(ZY, X)\right)\\
&+\sum_{i, j,k\neq (X_0 Y_0 Z_0)} \lambda_{i, j,k}(X Y Z) \left(C_{i, j,k}(X YZ) -  C_{i,j}(X Y, Z)- C_{i,k}(X Z, Y)- C_{k,j}(ZY, X) \right)+2 \ . 
\end{aligned}
\label{eq:derivecf3269}
\end{equation}
The terms inside the brackets are almost a cumulant expression, except the $+2$ term is missing, but one can add and subtract $2$ to obtain

\begin{equation}
\begin{aligned}
c(XYZ) &=\lambda_{X_0 Y_0 Z_0}(X YZ) c(X_0 Y_0 Z_0) + \sum_{i, j,k\neq (X_0 Y_0 Z_0)} \lambda_{i, j,k}(X Y Z) c(X_i Y_j Z_k) \\
&- 2\lambda_{X_0 Y_0 Z_0}(X YZ) - 2\sum_{i, j,k\neq (X_0 Y_0 Z_0)} \lambda_{i, j,k}(X Y Z) +2\\
&=\lambda_{X_0 Y_0 Z_0}(X YZ) c(X_0 Y_0 Z_0) + \sum_{i, j,k\neq (X_0 Y_0 Z_0)} \lambda_{i, j,k}(X Y Z) c(X_i Y_j Z_k) \\
&- 2\sum_{i, j,k} \lambda_{i, j,k}(X Y Z) +2\\
&=\lambda_{X_0 Y_0 Z_0}(X YZ) c(X_0 Y_0 Z_0) + \sum_{i, j,k\neq (X_0 Y_0 Z_0)} \lambda_{i, j,k}(X Y Z) c(X_i Y_j Z_k). 
\end{aligned}
\label{eq:derivecf32429}
\end{equation}

This is the final result  - the cumulant calculated using the measured correlation functions consists of the three correctly identified primary particle cumulant and the cumulant which consist of the rest of possible contributions. In such case, the correctly identified particle cumulant is
\begin{equation}
\begin{aligned}
c(X_0 Y_0 Z_0) &= \frac{1}{\lambda_{X_0 Y_0 Z_0}(X YZ)} \left(c(XYZ) - \sum_{i, j,k\neq (X_0 Y_0 Z_0)} \lambda_{i, j,k}(X Y Z) c(X_i Y_j Z_k) \right) \ .
\end{aligned}
\label{eq:derivecfBEAUTYapp}
\end{equation}

\newpage
\section{The ALICE Collaboration}
\label{app:collab}
% ALICE Collaboration author list for 2022-05-20
\begin{flushleft} 
\small

S.~Acharya\,\orcidlink{0000-0002-9213-5329}\,$^{\rm 125,132}$, 
D.~Adamov\'{a}\,\orcidlink{0000-0002-0504-7428}\,$^{\rm 86}$, 
A.~Adler$^{\rm 69}$, 
G.~Aglieri Rinella\,\orcidlink{0000-0002-9611-3696}\,$^{\rm 32}$, 
M.~Agnello\,\orcidlink{0000-0002-0760-5075}\,$^{\rm 29}$, 
N.~Agrawal\,\orcidlink{0000-0003-0348-9836}\,$^{\rm 50}$, 
Z.~Ahammed\,\orcidlink{0000-0001-5241-7412}\,$^{\rm 132}$, 
S.~Ahmad\,\orcidlink{0000-0003-0497-5705}\,$^{\rm 15}$, 
S.U.~Ahn\,\orcidlink{0000-0001-8847-489X}\,$^{\rm 70}$, 
I.~Ahuja\,\orcidlink{0000-0002-4417-1392}\,$^{\rm 37}$
% ...

\section*{Affiliation Notes}

$^{\rm I}$ Deceased\\
$^{\rm II}$ Also at: Italian National Agency for New Technologies, Energy and Sustainable Economic Development (ENEA), Bologna, Italy\\
$^{\rm III}$ Also at: Dipartimento DET del Politecnico di Torino, Turin, Italy\\
$^{\rm IV}$ Also at: Department of Applied Physics, Aligarh Muslim University, Aligarh, India\\
$^{\rm V}$ Also at: Institute of Theoretical Physics, University of Wroclaw, Poland\\
$^{\rm VI}$ Also at: University of Kansas, Lawrence, Kansas, United States\\
$^{\rm VII}$ Also at: An institution covered by a cooperation agreement with CERN\\

\section*{Collaboration Institutes}

$^{1}$ A.I. Alikhanyan National Science Laboratory (Yerevan Physics Institute) Foundation, Yerevan, Armenia\\
$^{2}$ AGH University of Science and Technology, Cracow, Poland\\
$^{3}$ Bogolyubov Institute for Theoretical Physics, National Academy of Sciences of Ukraine, Kiev, Ukraine\\
$^{4}$ Bose Institute, Department of Physics  and Centre for Astroparticle Physics and Space Science (CAPSS), Kolkata, India\\
$^{5}$ California Polytechnic State University, San Luis Obispo, California, United States\\
$^{6}$ Central China Normal University, Wuhan, China\\
$^{7}$ Centro de Aplicaciones Tecnol\'{o}gicas y Desarrollo Nuclear (CEADEN), Havana, Cuba\\
$^{8}$ Centro de Investigaci\'{o}n y de Estudios Avanzados (CINVESTAV), Mexico City and M\'{e}rida, Mexico\\
$^{9}$ Chicago State University, Chicago, Illinois, United States\\
$^{10}$ China Institute of Atomic Energy, Beijing, China\\
$^{11}$ Chungbuk National University, Cheongju, Republic of Korea\\
$^{12}$ Comenius University Bratislava, Faculty of Mathematics, Physics and Informatics, Bratislava, Slovak Republic\\
$^{13}$ COMSATS University Islamabad, Islamabad, Pakistan\\
$^{14}$ Creighton University, Omaha, Nebraska, United States\\
$^{15}$ Department of Physics, Aligarh Muslim University, Aligarh, India\\
$^{16}$ Department of Physics, Pusan National University, Pusan, Republic of Korea\\
$^{17}$ Department of Physics, Sejong University, Seoul, Republic of Korea\\
$^{18}$ Department of Physics, University of California, Berkeley, California, United States\\
$^{19}$ Department of Physics, University of Oslo, Oslo, Norway\\
$^{20}$ Department of Physics and Technology, University of Bergen, Bergen, Norway\\
$^{21}$ Dipartimento di Fisica, Universit\`{a} di Pavia, Pavia, Italy\\
$^{22}$ Dipartimento di Fisica dell'Universit\`{a} and Sezione INFN, Cagliari, Italy\\
$^{23}$ Dipartimento di Fisica dell'Universit\`{a} and Sezione INFN, Trieste, Italy\\
$^{24}$ Dipartimento di Fisica dell'Universit\`{a} and Sezione INFN, Turin, Italy\\
$^{25}$ Dipartimento di Fisica e Astronomia dell'Universit\`{a} and Sezione INFN, Bologna, Italy\\
$^{26}$ Dipartimento di Fisica e Astronomia dell'Universit\`{a} and Sezione INFN, Catania, Italy\\
$^{27}$ Dipartimento di Fisica e Astronomia dell'Universit\`{a} and Sezione INFN, Padova, Italy\\
$^{28}$ Dipartimento di Fisica `E.R.~Caianiello' dell'Universit\`{a} and Gruppo Collegato INFN, Salerno, Italy\\
$^{29}$ Dipartimento DISAT del Politecnico and Sezione INFN, Turin, Italy\\
$^{30}$ Dipartimento di Scienze MIFT, Universit\`{a} di Messina, Messina, Italy\\
$^{31}$ Dipartimento Interateneo di Fisica `M.~Merlin' and Sezione INFN, Bari, Italy\\
$^{32}$ European Organization for Nuclear Research (CERN), Geneva, Switzerland\\
$^{33}$ Faculty of Electrical Engineering, Mechanical Engineering and Naval Architecture, University of Split, Split, Croatia\\
$^{34}$ Faculty of Engineering and Science, Western Norway University of Applied Sciences, Bergen, Norway\\
$^{35}$ Faculty of Nuclear Sciences and Physical Engineering, Czech Technical University in Prague, Prague, Czech Republic\\
$^{36}$ Faculty of Physics, Sofia University, Sofia, Bulgaria\\
$^{37}$ Faculty of Science, P.J.~\v{S}af\'{a}rik University, Ko\v{s}ice, Slovak Republic\\
$^{38}$ Frankfurt Institute for Advanced Studies, Johann Wolfgang Goethe-Universit\"{a}t Frankfurt, Frankfurt, Germany\\
$^{39}$ Fudan University, Shanghai, China\\
$^{40}$ Gangneung-Wonju National University, Gangneung, Republic of Korea\\
$^{41}$ Gauhati University, Department of Physics, Guwahati, India\\
$^{42}$ Helmholtz-Institut f\"{u}r Strahlen- und Kernphysik, Rheinische Friedrich-Wilhelms-Universit\"{a}t Bonn, Bonn, Germany\\
$^{43}$ Helsinki Institute of Physics (HIP), Helsinki, Finland\\
$^{44}$ High Energy Physics Group,  Universidad Aut\'{o}noma de Puebla, Puebla, Mexico\\
$^{45}$ Horia Hulubei National Institute of Physics and Nuclear Engineering, Bucharest, Romania\\
$^{46}$ Indian Institute of Technology Bombay (IIT), Mumbai, India\\
$^{47}$ Indian Institute of Technology Indore, Indore, India\\
$^{48}$ INFN, Laboratori Nazionali di Frascati, Frascati, Italy\\
$^{49}$ INFN, Sezione di Bari, Bari, Italy\\
$^{50}$ INFN, Sezione di Bologna, Bologna, Italy\\
$^{51}$ INFN, Sezione di Cagliari, Cagliari, Italy\\
$^{52}$ INFN, Sezione di Catania, Catania, Italy\\
$^{53}$ INFN, Sezione di Padova, Padova, Italy\\
$^{54}$ INFN, Sezione di Pavia, Pavia, Italy\\
$^{55}$ INFN, Sezione di Torino, Turin, Italy\\
$^{56}$ INFN, Sezione di Trieste, Trieste, Italy\\
$^{57}$ Inha University, Incheon, Republic of Korea\\
$^{58}$ Institute for Gravitational and Subatomic Physics (GRASP), Utrecht University/Nikhef, Utrecht, Netherlands\\
$^{59}$ Institute of Experimental Physics, Slovak Academy of Sciences, Ko\v{s}ice, Slovak Republic\\
$^{60}$ Institute of Physics, Homi Bhabha National Institute, Bhubaneswar, India\\
$^{61}$ Institute of Physics of the Czech Academy of Sciences, Prague, Czech Republic\\
$^{62}$ Institute of Space Science (ISS), Bucharest, Romania\\
$^{63}$ Institut f\"{u}r Kernphysik, Johann Wolfgang Goethe-Universit\"{a}t Frankfurt, Frankfurt, Germany\\
$^{64}$ Instituto de Ciencias Nucleares, Universidad Nacional Aut\'{o}noma de M\'{e}xico, Mexico City, Mexico\\
$^{65}$ Instituto de F\'{i}sica, Universidade Federal do Rio Grande do Sul (UFRGS), Porto Alegre, Brazil\\
$^{66}$ Instituto de F\'{\i}sica, Universidad Nacional Aut\'{o}noma de M\'{e}xico, Mexico City, Mexico\\
$^{67}$ iThemba LABS, National Research Foundation, Somerset West, South Africa\\
$^{68}$ Jeonbuk National University, Jeonju, Republic of Korea\\
$^{69}$ Johann-Wolfgang-Goethe Universit\"{a}t Frankfurt Institut f\"{u}r Informatik, Fachbereich Informatik und Mathematik, Frankfurt, Germany\\
$^{70}$ Korea Institute of Science and Technology Information, Daejeon, Republic of Korea\\
$^{71}$ KTO Karatay University, Konya, Turkey\\
$^{72}$ Laboratoire de Physique des 2 Infinis, Ir\`{e}ne Joliot-Curie, Orsay, France\\
$^{73}$ Laboratoire de Physique Subatomique et de Cosmologie, Universit\'{e} Grenoble-Alpes, CNRS-IN2P3, Grenoble, France\\
$^{74}$ Lawrence Berkeley National Laboratory, Berkeley, California, United States\\
$^{75}$ Lund University Department of Physics, Division of Particle Physics, Lund, Sweden\\
$^{76}$ Nagasaki Institute of Applied Science, Nagasaki, Japan\\
$^{77}$ Nara Women{'}s University (NWU), Nara, Japan\\
$^{78}$ National and Kapodistrian University of Athens, School of Science, Department of Physics , Athens, Greece\\
$^{79}$ National Centre for Nuclear Research, Warsaw, Poland\\
$^{80}$ National Institute of Science Education and Research, Homi Bhabha National Institute, Jatni, India\\
$^{81}$ National Nuclear Research Center, Baku, Azerbaijan\\
$^{82}$ National Research and Innovation Agency - BRIN, Jakarta, Indonesia\\
$^{83}$ Niels Bohr Institute, University of Copenhagen, Copenhagen, Denmark\\
$^{84}$ Nikhef, National institute for subatomic physics, Amsterdam, Netherlands\\
$^{85}$ Nuclear Physics Group, STFC Daresbury Laboratory, Daresbury, United Kingdom\\
$^{86}$ Nuclear Physics Institute of the Czech Academy of Sciences, Husinec-\v{R}e\v{z}, Czech Republic\\
$^{87}$ Oak Ridge National Laboratory, Oak Ridge, Tennessee, United States\\
$^{88}$ Ohio State University, Columbus, Ohio, United States\\
$^{89}$ Physics department, Faculty of science, University of Zagreb, Zagreb, Croatia\\
$^{90}$ Physics Department, Panjab University, Chandigarh, India\\
$^{91}$ Physics Department, University of Jammu, Jammu, India\\
$^{92}$ Physics Department, University of Rajasthan, Jaipur, India\\
$^{93}$ Physics Program and International Institute for Sustainability with Knotted Chiral Meta Matter (SKCM2), Hiroshima University, Hiroshima, Japan\\
$^{94}$ Physikalisches Institut, Eberhard-Karls-Universit\"{a}t T\"{u}bingen, T\"{u}bingen, Germany\\
$^{95}$ Physikalisches Institut, Ruprecht-Karls-Universit\"{a}t Heidelberg, Heidelberg, Germany\\
$^{96}$ Physik Department, Technische Universit\"{a}t M\"{u}nchen, Munich, Germany\\
$^{97}$ Politecnico di Bari and Sezione INFN, Bari, Italy\\
$^{98}$ Research Division and ExtreMe Matter Institute EMMI, GSI Helmholtzzentrum f\"ur Schwerionenforschung GmbH, Darmstadt, Germany\\
$^{99}$ Saga University, Saga, Japan\\
$^{100}$ Saha Institute of Nuclear Physics, Homi Bhabha National Institute, Kolkata, India\\
$^{101}$ School of Physics and Astronomy, University of Birmingham, Birmingham, United Kingdom\\
$^{102}$ Secci\'{o}n F\'{\i}sica, Departamento de Ciencias, Pontificia Universidad Cat\'{o}lica del Per\'{u}, Lima, Peru\\
$^{103}$ Stefan Meyer Institut f\"{u}r Subatomare Physik (SMI), Vienna, Austria\\
$^{104}$ SUBATECH, IMT Atlantique, Nantes Universit\'{e}, CNRS-IN2P3, Nantes, France\\
$^{105}$ Suranaree University of Technology, Nakhon Ratchasima, Thailand\\
$^{106}$ Technical University of Ko\v{s}ice, Ko\v{s}ice, Slovak Republic\\
$^{107}$ The Henryk Niewodniczanski Institute of Nuclear Physics, Polish Academy of Sciences, Cracow, Poland\\
$^{108}$ The University of Texas at Austin, Austin, Texas, United States\\
$^{109}$ Universidad Aut\'{o}noma de Sinaloa, Culiac\'{a}n, Mexico\\
$^{110}$ Universidade de S\~{a}o Paulo (USP), S\~{a}o Paulo, Brazil\\
$^{111}$ Universidade Estadual de Campinas (UNICAMP), Campinas, Brazil\\
$^{112}$ Universidade Federal do ABC, Santo Andre, Brazil\\
$^{113}$ University of Cape Town, Cape Town, South Africa\\
$^{114}$ University of Houston, Houston, Texas, United States\\
$^{115}$ University of Jyv\"{a}skyl\"{a}, Jyv\"{a}skyl\"{a}, Finland\\
$^{116}$ University of Kansas, Lawrence, Kansas, United States\\
$^{117}$ University of Liverpool, Liverpool, United Kingdom\\
$^{118}$ University of Science and Technology of China, Hefei, China\\
$^{119}$ University of South-Eastern Norway, Kongsberg, Norway\\
$^{120}$ University of Tennessee, Knoxville, Tennessee, United States\\
$^{121}$ University of the Witwatersrand, Johannesburg, South Africa\\
$^{122}$ University of Tokyo, Tokyo, Japan\\
$^{123}$ University of Tsukuba, Tsukuba, Japan\\
$^{124}$ University Politehnica of Bucharest, Bucharest, Romania\\
$^{125}$ Universit\'{e} Clermont Auvergne, CNRS/IN2P3, LPC, Clermont-Ferrand, France\\
$^{126}$ Universit\'{e} de Lyon, CNRS/IN2P3, Institut de Physique des 2 Infinis de Lyon, Lyon, France\\
$^{127}$ Universit\'{e} de Strasbourg, CNRS, IPHC UMR 7178, F-67000 Strasbourg, France, Strasbourg, France\\
$^{128}$ Universit\'{e} Paris-Saclay Centre d'Etudes de Saclay (CEA), IRFU, D\'{e}partment de Physique Nucl\'{e}aire (DPhN), Saclay, France\\
$^{129}$ Universit\`{a} degli Studi di Foggia, Foggia, Italy\\
$^{130}$ Universit\`{a} del Piemonte Orientale, Vercelli, Italy\\
$^{131}$ Universit\`{a} di Brescia, Brescia, Italy\\
$^{132}$ Variable Energy Cyclotron Centre, Homi Bhabha National Institute, Kolkata, India\\
$^{133}$ Warsaw University of Technology, Warsaw, Poland\\
$^{134}$ Wayne State University, Detroit, Michigan, United States\\
$^{135}$ Westf\"{a}lische Wilhelms-Universit\"{a}t M\"{u}nster, Institut f\"{u}r Kernphysik, M\"{u}nster, Germany\\
$^{136}$ Wigner Research Centre for Physics, Budapest, Hungary\\
$^{137}$ Yale University, New Haven, Connecticut, United States\\
$^{138}$ Yonsei University, Seoul, Republic of Korea\\
$^{139}$  Zentrum  f\"{u}r Technologie und Transfer (ZTT), Worms, Germany\\
$^{140}$ Affiliated with an institute covered by a cooperation agreement with CERN\\
$^{141}$ Affiliated with an international laboratory covered by a cooperation agreement with CERN\\

\end{flushleft} 
  %%%%%%% done by webmaster team
%TC:endignore
\end{document}